\documentclass[a4paper,11pt]{article}

\pdfoutput=1

\usepackage[utf8]{inputenc}
\usepackage{amsmath}
\usepackage{amssymb}
\usepackage{amsfonts}
\usepackage{mathrsfs}
\usepackage{graphicx}
\usepackage{booktabs,adjustbox}
\usepackage{caption}
\usepackage{slashed}
\usepackage{verbatim}
\usepackage{float}
\usepackage{setspace}
\usepackage{subfig}
\usepackage{jheppub}

\usepackage{color}
\usepackage{ulem}

\usepackage{bookmark}
\usepackage{wasysym}

\makeatletter
\newcommand{\thickhline}{%
	\noalign {\ifnum 0=`}\fi \hrule height 1pt
	\futurelet \reserved@a \@xhline
}
\makeatother

\usepackage[utf8]{inputenc}
\usepackage{times}
\usepackage{float}

\usepackage{setspace}

\title{\boldmath Forbidden neutrinogenesis}

\author[a]{Shinya Kanemura}
\author[a]{and Shao-Ping Li}

\affiliation[a]{Department of Physics, The  University of Osaka, Toyonaka, Osaka 560-0043, Japan}

\emailAdd{kanemu@het.phys.sci.osaka-u.ac.jp}
\emailAdd{lisp@het.phys.sci.osaka-u.ac.jp}

 \preprint{OU-HET 1256} 
 
\abstract{ The origin of neutrino masses can  be simply attributed to a new scalar beyond the Standard Model.  We demonstrate that  leptogenesis can    explain the baryon asymmetry of the universe already  in such a minimal framework.  Different from traditional leptogenesis,   the realization here exploits the thermal behavior of leptons at finite temperatures, which is otherwise kinematically  forbidden in vacuum. We present   detailed calculations of the CP asymmetry  in the Schwinger-Keldysh Closed-Time-Path formalism,  and compute the   asymmetry evolution  via the Kadanoff-Baym equation. Such minimal   forbidden neutrinogenesis establishes a direct link between the baryon asymmetry  and the CP-violating phase from neutrino mixing,  making the scenario a compelling target   in neutrino oscillation experiments. Complementary probes from cosmology, flavor physics and colliders are also briefly   discussed. 
	
}

\begin{document}
	\maketitle
	\flushbottom
	
	\section{Introduction}
	\label{sec:intro}
Baryogenesis via leptogenesis~\cite{Fukugita:1986hr,Luty:1992un,Covi:1996wh} is a simple mechanism that can explain the baryon asymmetry of the universe (BAU), which    typically  occurs   above the sphaleron decoupling temperature $T_{\rm sph}\approx 132$~GeV~\cite{Kuzmin:1985mm,DOnofrio:2014rug}. 
While the study of leptogenesis is mostly based on  perturbative lepton-number violation~\cite{Giudice:2003jh,Buchmuller:2004nz,Davidson:2008bu,Fong:2012buy}, leptogenesis can also be realized with a total lepton number  conserved and  shared between the Standard Model (SM) and a hidden sector.  Such lepton-number conserving leptogenesis can also feature  connections to the neutrino mass origin when  the hidden sector consists of  light Majorana or Dirac   right-handed neutrinos~\cite{Akhmedov:1998qx,Dick:1999je}, and links to dark matter cogenesis~\cite{Cheung:2011if,Kanemura:2024dqv}.

Leptogenesis  is a high-temperature process,  where  finite-temperature corrections should   generally contribute to the generation of CP asymmetries. Theoretical development of leptogenesis in recent years confirms this expectation, and has brought us an interesting phenomenon: the CP asymmetry at high temperatures  can be induced by  a pure plasma effect  that would be  otherwise kinematically  forbidden  in vacuum. Generally, such a kind of \textit{forbidden leptogenesis} can be  exploited by using nonthermal quantum field theory, or the  Schwinger-Keldysh Closed-Time-Path (SK-CTP) formalism~\cite{Chou:1984es,Calzetta:1986cq,Berges:2004yj}, as  mostly studied in  lepton-number violating scenarios~\cite{Garny:2009qn,Beneke:2010wd,Garbrecht:2010sz,Garny:2010nj,Beneke:2010dz,Drewes:2012ma,Garbrecht:2011aw,Garny:2011hg,BhupalDev:2014oar,Frossard:2012pc,Garbrecht:2012qv,Hambye:2016sby}. 
	
Based on the SK-CTP formalism,  the evolution of CP asymmetries   is determined by the Kadanoff-Baym (KB) equation, which is  more technically and computationally challenging than the conventional Boltzmann equation.  However,   the KB   equation does not suffer from the double-counting issue found in the Boltzmann equation, which is related to the real-intermediate-state subtraction~\cite{Kolb:1979qa,Giudice:2003jh}. Instead,  the KB collision rates automatically include all the relevant processes from  two-loop diagrams, where the  subtraction of on-shell scattering is guaranteed~\cite{Beneke:2010wd,Kanemura:2024dqv}. 

Regarding the complexity of  finite-temperature CP asymmetries, it is recently   pointed out in Ref.~\cite{Kanemura:2024dqv} that if the total lepton number is (nearly) conserved and shared between the SM leptons and hidden particles, the origin of CP asymmetries in forbidden leptogenesis will become   easy to trace.  In that regime, one can use perturbative lepton-number conservation to calculate the hidden  asymmetries, where a resonant enhancement from soft-lepton resummation clearly shows up in   two-loop self-energy diagrams.  In contrast to resonant leptogenesis via vacuum mass degeneracy~\cite{Pilaftsis:2003gt,Pilaftsis:2005rv}, the appearance of soft-lepton resonance is itself a SM prediction and the resonant enhancement  is protected from finite width under perturbative finite-temperature field theory.  

 A compelling feature of forbidden leptogenesis is that CP asymmetries can   be generated with  minimal particle physics,  where traditional leptogenesis   based on  vacuum quantum field theory cannot be realized. Consequently,  forbidden leptogenesis can open more channels to explain the BAU without invoking abundant particle content. This comes from the expectation that finite-temperature corrections induce more absorptive contributions (kinetic phases)  than from vacuum   loop diagrams. 
 It also implies that in  non-minimal particle-physics scenarios considered insofar,  forbidden leptogenesis can readily    contribute an irreducible CP asymmetry at high temperatures, which in some cases could even dominate over the conventional CP asymmetry and hence should be considered consistently whenever traditional  leptogenesis is  at work. 

In  this paper, we consider   forbidden leptogenesis in a minimal framework, where   a new neutrinophilic scalar is introduced to the SM.  Concretely, we consider a two-Higgs-doublet model with right-handed neutrinos,  in which the vacuum expectation value of the second Higgs doublet gives   Dirac masses to SM neutrinos~\cite{Gabriel:2006ns,Davidson:2009ha}. 
Both right-handed neutrinos and their asymmetries are co-generated via    neutrinophilic scalar decay, and hence is dubbed \textit{forbidden neutrinogenesis}.  The total lepton number is conserved and shared between   the SM lepton and  right-handed neutrino sectors. 
 In  such a minimal neutrinophilic scalar scenario,   the CP violation for the BAU is   sourced by  the   Pontecorvo–Maki–Nakagawa–Sakata (PMNS) mixing matrix~\cite{Pontecorvo:1957qd,Maki:1962mu}. In particular, the Dirac CP-violating phase in the PMNS matrix determines the sign of the baryon asymmetry, making this scenario  highly falsifiable  in terms of neutrino oscillation experiments.

We begin in section~\ref{sec:framework} with the introduction of the  neutrinophilic scalar scenario. Section~\ref{sec:th-cal} is devoted to elaborating  the calculations of forbidden neutrinogenesis in the SK-CTP formalism with the KB equation, where the  resonant enhancement from soft-lepton resummation can be traced.  In section~\ref{sec:KBvsBol}, we will make a comparison between the KB and Boltzmann equations, pointing out some issues in previous work. In section~\ref{sec:pheno}, we will briefly  discuss some phenomenological signals, showing the complementary probes in cosmology, flavor physics and colliders.  Finally,   we will present the conclusions in section~\ref{sec:con}.  The propagators and resummation in the SK-CTP formalism, as well as some technical calculations of finite-temperature CP asymmetries used  in section~\ref{sec:th-cal} will be  relegated to some  appendices.

\section{A minimal neutrinophilic scalar scenario}\label{sec:framework}

One of the simplest explanations for the SM neutrino masses is to introduce a second Higgs doublet beyond the SM~\cite{Gabriel:2006ns,Davidson:2009ha}.
The  scalar potential featuring  an approximate  global $U(1)$ symmetry reads
\begin{align}\label{eq:pot}
	V(\phi_{1},\phi_2)&=\mu_{1}^2 \phi_{1}^\dagger \phi_{1}+\mu_{2}^2 \phi_2^\dagger \phi_2-\left[\mu^2 \phi_{1}^\dagger \phi_2+\text{h.c.}\right]
	\nonumber\\[0.2cm]
	&+\frac{\lambda_1}{2}(\phi_{1}^\dagger \phi_{1})^2+\frac{\lambda_2}{2}(\phi_2^\dagger \phi_2)^2+\lambda_3(\phi_{1}^\dagger\phi_{1})(\phi_2^\dagger \phi_2)+\lambda_4 (\phi_{1}^\dagger \phi_2)(\phi_2^\dagger \phi_{1}) ,
\end{align}
where the $\mu^2$ term denotes a soft symmetry-breaking source~\cite{Davidson:2009ha,Aoki:2009ha,Branco:2011iw}. It should be mentioned  that the scalar potential may also be constructed with an approximate $Z_2$ symmetry, where an additional quartic term $(\phi_{1}^\dagger \phi_2)^2$ can arise.  This additional  quartic term may  become relevant e.g., when considering collider phenomenology associated with the scalar sector.  For simplicity, we will not consider this quartic term, which is   irrelevant to leptogenesis discussed in this paper. 

 The $\mu^2$ term  can be identified  as a low-scale effective interaction, which may be induced via a super-renormalizable portal to inflaton field $\phi_{\rm inf}$ through $\mu_{\rm UV} (\phi_{1}^\dagger \phi_2) \phi_{\rm inf}$, with $\mu_{\rm UV}$ a   dimensionful coupling at high scales. This super-renormalizable portal could open interesting connections to high-scale physics, such as  spontaneous symmetry breaking and the associated  topological defects.  In this paper, we will turn agnostic on the origin of this symmetry-breaking term, and simply take $\mu$ as a free parameter. 
 The electroweak gauge symmetry breaking is considered to follow the development of  nonzero vacuum expectation values  owing to a negative $\mu_1^2$. On the other hand, we will consider a positive  $\mu_2^2$ to avoid very  light pseudo Nambu-Goldstone (NG) bosons~\cite{Davidson:2009ha}. As will be discussed in forbidden leptogenesis, the situation   $\mu_2^2>0$  provides a vacuum mass   for the second Higgs $\phi_2$, and  helps to drive $\phi_2$ into the nonthermal regime in the early universe, consequently assisting the generation of CP asymmetry.

Right-handed neutrinos would  exclusively couple to the second Higgs doublet $\phi_2$ if they are also charged under the global $U(1)$  symmetry. In addition,  $\phi_2$   coupling to the SM quarks and right-handed charged-lepton singlets will also be absent. 
Consequently, the new physics Yukawa interaction comes from the   neutrino coupling, which   is built upon  the following lepton portal~\cite{Gabriel:2006ns,Davidson:2009ha}:
\begin{align}\label{eq:lag}
	\mathcal{L}= -y_{i\alpha} \bar \ell_i \tilde{\phi}_2 \nu_{R\alpha}+\rm h.c.\,,
\end{align}
where $\nu_{R\alpha}$ are the three right-handed Dirac counterparts of the SM left-handed neutrinos, and $\tilde{\phi}_2\equiv i \sigma_2 \phi_2^*$ with $\sigma_2$ the second Pauli matrix.   For Dirac neutrinos, there is also a global  lepton-number $U(1)_L$ symmetry, such that  the total lepton number is  perturbatively  conserved, and the  Majorana mass term $m\overline{\nu_R^c}\nu_R$ is forbidden.

The Higgs doublets from Eq.~\eqref{eq:pot} can be parameterized as
\begin{align}
	\phi_{a}=\left( \begin{array}{cc}  \varphi_a^+ \\ \dfrac{1}{\sqrt{2}}(v_a+\rho_a+i\eta_a)
	\end{array}
	\right), \quad a=1,2\,, 
	\label{eq:vev}
\end{align}
where the vacuum expectation values satisfy $(v_1^2+v_2^2)^{1/2}=246$~GeV. In   the limit of $v_1\gg v_2, \mu$,  the tadpole equations yield
\begin{align}
v_1 \approx \sqrt{-\frac{2\mu_1^2 }{\lambda_1}}\,, 
\qquad
 v_2\approx \frac{2\mu^2 v_1}{2\mu_2^2+ \lambda_{34} v_1^2}\,,
\end{align}
with    $\lambda_{34}\equiv \lambda_3+\lambda_4$.
The mixed charged components  will give rise to   charged NG bosons ($G^\pm$) and   physical charged scalars ($H^\pm$), while the mixed $\eta_a$ fields lead to one neutral NG boson ($G^0$) and one pseudoscalar ($A$). The transformation rules read
\begin{align}
	\left( \begin{array}{cc}  \varphi_1^\pm \\ \varphi_2^\pm
	\end{array}
	\right)=\left( \begin{array}{cc} \cos \beta & -\sin \beta \\ \sin \beta & \cos \beta
	\end{array}  \right)  \left( \begin{array}{cc}  G^\pm \\ H^\pm
	\end{array}
	\right), \qquad   	\left( \begin{array}{cc}  \eta_1 \\ \eta_2
	\end{array}
	\right)=\left( \begin{array}{cc} \cos \beta & -\sin \beta \\ \sin \beta & \cos \beta
	\end{array}  \right)\left( \begin{array}{cc}  G^0 \\ A
	\end{array}
	\right),	\label{eq:beta}
\end{align}
where the mixing angle is given by 
\begin{align}
\tan\beta=\frac{ v_2}{v_1}\,.
\end{align}
On the other hand, the mixed  $\rho_a$ fields  give rise to the SM Higgs boson ($h$) and a new scalar boson ($H$), with the transformation defined by
\begin{align}
	\left( \begin{array}{cc}  \rho_1 \\ \rho_2
		\end{array}
		\right)=\left( \begin{array}{cc} \cos \alpha & -\sin \alpha \\ \sin \alpha & \cos \alpha
		\end{array}  \right)\left( \begin{array}{cc}  h \\ H
		\end{array}
		\right).	\label{eq:alpha}
\end{align}
The mixing angle $\alpha$ is found to be 
\begin{align}
	\tan(2\alpha)&=\frac{4(\mu^2-\lambda_{34} v_1 v_2)}{(-3\lambda_1+\lambda_{34})v_1^2-2\mu_1^2+2\mu_2^2+3\lambda_2 v_2^2- \lambda_{34}v_2^2}
	\\[0.25cm]
	&\approx \frac{2v_2(2\mu_2^2/v_1- \lambda_{34} v_1)}{(-2\lambda_1+ \lambda_{34})v_1^2 +2\mu_2^2}\,,
\end{align}
where the second approximation is derived in the limit of  $v_1\gg v_2,\mu$. 
For $\mu_2\gtrsim v_1$, we arrive at
\begin{align}
 \tan(2\alpha)\simeq \mathcal{O}(v_2/v_1)\,, 
\end{align}
indicating  that   mixing of $h$ and $H$ is suppressed and $\phi_1$ will reduce to the SM Higgs doublet (denoted by $\phi_{\rm SM}$) while $\phi_2$ becomes a neutrinophilic Higgs doublet (denoted by $\phi$), 
\begin{align}\label{eq:align}
	\phi_1\approx \phi_{\rm SM}\,, \quad \phi_2\approx \phi\,.
\end{align}
After gauge symmetry breaking,   SM neutrinos acquire masses from $\phi$, with
\begin{align}\label{eq:nu-mass}
	m=\frac{y}{\sqrt{2}}v_2\,.
\end{align}

In the following analysis, we will take the   alignment limit given by Eq.~\eqref{eq:align}, which will be    justified   by   $v_2\ll v_1$   inferred from forbidden neutrinogenesis. 
In the  limit  of Eq.~\eqref{eq:align}, 
the mass spectrum of the Higgs bosons reads
\begin{align}\label{eq:Higgsmass}
m_h^2\approx \lambda_1 v_1^2\,, \quad m_H^2= m_A^2\approx \mu_2^2+\frac{\lambda_{34}}{2} v_1^2\,, \quad m^2_{H^\pm}\approx\mu_2^2+\frac{ \lambda_3}{2} v_1^2\,.
\end{align}
The $\lambda$ parameters in Eq.~\eqref{eq:pot} are theoretically constrained by the requirements of vacuum stability~\cite{Sher:1988mj,Nie:1998yn,Kanemura:1999xf,Branco:2011iw}, triviality~\cite{Wilson:1971dh,Inoue:1982ej,Dashen:1983ts,Callaway:1983zd,Luscher:1988uq}, and also by perturbative unitarity~\cite{Kanemura:1993hm,Akeroyd:2000wc,Ginzburg:2005dt,Grinstein:2015rtl}.  In addition, there are  experimental constraints on the Higgs mass spectrum. In particular, the electroweak corrections, which are commonly parameterized by  the oblique parameters~\cite{Toussaint:1978zm,Bertolini:1985ia,Peskin:1990zt,Peskin:1991sw}, require a quasi-degenerate mass between   $H^\pm$ and $A/H$~\cite{Gerard:2007kn,Haber:2010bw,Kanemura:2011sj}, which is guaranteed if $\lambda_4$ is small.  Direct searches from the LEP collaborations have also excluded a charged-scalar mass below 80~GeV~\cite{ALEPH:2013htx}.  These constraints will be considered  in the following analysis, which helps to determine whether     the forbidden neutrinogenesis can be realized in the minimal neutrinophilic scalar scenario.  

Finally, it is worth highlighting   the aforementioned structure of the neutrinophilic two-Higgs-doublet model in some aspects.  Although the approximate $U(1)$ or $Z_2$ symmetry may not be mandatory by nature, it makes the Yukawa interaction~\eqref{eq:lag} a minimal and distinguishable scenario from the traditional two-Higgs-doublet models (see Ref.~\cite{Branco:2011iw} for a review). 
Without the approximate symmetry,  right-handed neutrinos would also   couple to  the SM Higgs doublet having   $v_1\approx 246$~GeV, 
but such interactions are experimentally less interesting and irrelevant due to the exceedingly weak Yukawa couplings $\lesssim \mathcal{O}(10^{-13})$,  even though the feebleness of neutrino Yukawa couplings is technically natural on the theoretical side.  This is one of the motivations for introducing the second Higgs doublet with the approximate  symmetry, where $v_2$   can be much smaller  than $v_1$  and hence the  neutrino Yukawa couplings can be  larger by orders of magnitude.  

The soft-breaking $\mu^2$ term helps to induce a nonzero vacuum expectation value  in the second Higgs doublet after the electroweak gauge symmetry is broken, and in addition, 
helps to lower the $v_2$ scale such that tiny neutrino masses can be generated without feeble Yukawa couplings.  On the other hand, the smallness of $\mu$ is technically natural in the sense that it is the symmetry breaking source~\cite{Davidson:2009ha}. As will be shown later, a ratio $v_2/v_1=\mathcal{O}(10^{-11})$ is favored to realize the minimal forbidden neutrinogenesis. In particular, such a  small ratio  ensures that tree-level couplings between the new scalars ($H^\pm, H, A$) and SM quarks/right-handed charged leptons, as well as   loop-induced flavor-changing neutral currents, such as $h, Z\to \ell_\alpha \ell_\beta$, are strongly suppressed~\cite{Bertuzzo:2015ada}. 

 \section{Resonant forbidden neutrinogenesis}\label{sec:th-cal} 
Due to lepton-number conservation,  the SM lepton asymmetries generated via the Yukawa interaction~\eqref{eq:lag} will be accompanied by right-handed neutrino asymmetries. As long as the asymmetry in the right-handed neutrino sector is maintained by the out-of-equilibrium condition persisted prior to   sphaleron decoupling,  there would also be a  net asymmetry in the SM lepton sector. The SM lepton asymmetry will be converted into the baryon asymmetry via active sphaleron processes, while the asymmetry in the right-handed neutrino sector     accumulates over time.  Even though the SM leptons are in quasi-thermal equilibrium,    the  net SM lepton asymmetry would not be washed out before sphaleron processing. Instead, it  will be redistributed among   lepton flavors.  

The final baryon asymmetry   is simply  determined by the amount of asymmetries stored in the $\nu_R$ sector. Based on the sphaleron conversion efficiency, chemical equilibrium and  perturbative lepton-number conservation, one can obtain the relation between the  baryon and right-handed neutrino  asymmetries~\cite{Harvey:1990qw}:
\begin{align}\label{eq:YB}
	Y_{B}\equiv \frac{n_{B}- n_{\bar B}}{s_{\rm SM}}\approx 0.35 \sum_\alpha Y_{\nu_{R\alpha}}\,,
\end{align}	 
where $Y_{\nu_{R\alpha}} \equiv (n_{\nu_{R\alpha}}- n_{\bar \nu_{R\alpha}})/s_{\rm SM}$  denotes the $\nu_R$ asymmetry  of flavor $\alpha$ normalized to     the SM entropy density,
\begin{align}\label{eq:entropy}
	s_{\rm SM}=g_s(T) \frac{2\pi^2}{45}T^3\,,
\end{align}
with  $g_s(T)$   the effective degrees of freedom. Typically, we have   $g_s(T)\approx 106.75$ during leptogenesis.     The total $\nu_R$ asymmetry should  match  the observed $Y_B$ at present day~\cite{Planck:2018vyg}:
   \begin{align}\label{eq:YB-exp}
Y_B\approx 8.75\times 10^{-11}\,.
   \end{align}
  In addition, one can circumvent  the dynamics of asymmetry redistribution in   SM lepton flavors, using  the evolution of the  accompanying $\nu_R$ asymmetry  as a result of perturbative lepton-number conservation. This technical  treatment will allow us to visualize the role of resummed thermal leptons in contributing to finite-temperature  CP asymmetries.

\subsection{Kadanoff-Baym equation}
To calculate the CP asymmetry in the $\nu_R$ sector, we start from the   KB kinetic equation for right-handed Dirac neutrinos~\cite{Prokopec:2003pj,Prokopec:2004ic,Beneke:2010wd,Kanemura:2024dqv}:
\begin{align}\label{eq:0th-KB}
	\gamma^0 \frac{d}{d t} (i\slashed S_{\nu_\alpha}^{\lessgtr})=(-i\slashed\Sigma_{\nu_\alpha}^{>})(i\slashed S_{\nu_\alpha}^{<})-(-i\slashed \Sigma_{\nu_\alpha}^{<})(i\slashed S_{\nu_\alpha}^>)\,,
\end{align}
where $i \slashed S_{\nu_\alpha}^{\lessgtr}$ denote the Wightman functions and $-i\slashed\Sigma_{\nu_{\alpha}}^{\gtrless}$ the self-energy amplitudes for  $\nu_{R}$ of flavor $\alpha$, with   the slashed symbol   highlighting the spinor structures of  fermion self-energy amplitudes and propagators. Notice that  $d(i \slashed S_{\nu_\alpha}^{<})/dt$ and $d(i \slashed S_{\nu_\alpha}^{>})/dt$ have the same rate from the right-hand side. 

The particle-number asymmetry of right-handed neutrinos, $\Delta n_\alpha \equiv n_{\nu_\alpha}- n_{\bar \nu_\alpha}$, can be obtained from Eq.~\eqref{eq:0th-KB} by integrating over the 4-momentum of neutrinos and performing the Dirac trace, which gives  rise to 
\begin{align}\label{eq:KB}
	\frac{d\Delta n_\alpha}{d t}=\frac{1}{2}\int_p \text{Tr}\left[i\slashed\Sigma_{\nu_\alpha}^> i\slashed S_{\nu_\alpha}^<-i\slashed\Sigma_{\nu_\alpha}^<i \slashed S_{\nu_\alpha}^>\right],
\end{align}
with $\rm Tr$ being  the Dirac trace and   
\begin{align}
	\int_p \equiv\int \frac{d^4 p}{(2\pi)^4}\,,
\end{align} 
for shorthand.
The KB collision rates on the right-hand side of Eq.~\eqref{eq:KB} contain  the washout effect  at one-loop level and the CP-violating source  at two-loop level.

\subsection{One-loop washout rate}\label{sec:washout}

\begin{figure}[t]
	\centering
	\includegraphics[scale=0.7]{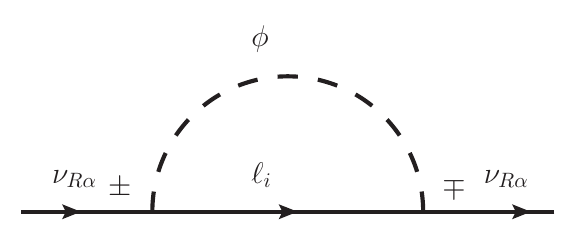} 
	\caption{\label{fig:washout} The one-loop self-energy diagram of $\nu_{R\alpha}$ contributing to the washout rate.  The   thermal indices $+-$ and $-+$ correspond  to the amplitudes $\slashed{\Sigma}_{\nu_\alpha}^{<}$ and $\slashed{\Sigma}_{\nu_\alpha}^{>}$, respectively.  }
\end{figure}

The washout processes are determined by  the one-loop self-energy diagrams of $\nu_\alpha$, as shown in Fig.~\ref{fig:washout}.   The one-loop amplitudes $\slashed \Sigma_{\nu_\alpha}^{\lessgtr}$ read\footnote{An easy way to remember the correspondence between the thermal indices $\pm$ and the symbols $\lessgtr$ is that   the time variable for  $+$ in the SK-CTP formalism is always earlier (smaller $<$) than for $-$.}
\begin{align}\label{eq:S<_chi2}
	i\slashed{\Sigma}_{\nu_\alpha}^{+-}(p_{})&\equiv 	i\slashed{\Sigma}_{\nu_\alpha}^<(p_{})=2|y_{\alpha}|^2\int_{p_\ell} \int_{p_\phi} (2\pi)^4\delta^4(p_{}-p_\ell+p_\phi)P_L i\slashed{{S}}^<_{\ell}P_R iG_\phi^>\,,
	\\[0.3cm]
	i\slashed{\Sigma}_{\nu_\alpha}^{-+}(p_{})&\equiv 	i\slashed{\Sigma}_{\nu_\alpha}^>(p_{})=2|y_{\alpha}|^2\int_{p_\ell} \int_{p_\phi} (2\pi)^4\delta^4(p_{}-p_\ell+p_\phi)P_L i\slashed{{S}}^>_{\ell}P_R iG_\phi^<\,,\label{eq:S>_chi3}
\end{align}
where  $(+-), (-+)$ denote the thermal indices, and the factor of 2 comes from gauge $SU(2)_L$ degeneracy. The Yukawa coupling is defined as
\begin{align}
	|y_{\alpha}|^2\equiv \sum_{i=e,\mu,\tau}y_{i\alpha}y^*_{i\alpha}\,,
\end{align}
 and $\slashed{{S}}^{\lessgtr}_{\ell}, G_\phi^{\gtrless}$ denote the   lepton and scalar  propagators  collected  in Appendix~\ref{append:CTP-1}.
 
Calculating the one-loop self-energy amplitudes straightforwardly, we obtain the washout rate:
\begin{align}
	\mathcal{W}&=\frac{1}{2}\int_p \text{Tr}\left[i\slashed\Sigma_{\nu_\alpha}^> i\slashed S_{\nu_\alpha}^<-i\slashed\Sigma_{\nu_\alpha}^<i \slashed S_{\nu_\alpha}^>\right]_{\rm 1-loop}
	\nonumber \\[0.2cm]
	&=	-\frac{|y_\alpha|^2m_\phi^2}{32\pi^3}\int_0^\infty dp \int_{m_\phi^2/(4p)}^\infty dp_\ell \left[f_\alpha(p)-\bar f_\alpha(p)\right]\left[f^{\rm eq}_\phi(p_\ell+p)+f^{\rm eq}_\ell(p_\ell)\right],\label{eq:W-trace2}
\end{align}
where we have used $\text{Tr}[P_L \slashed{p}_\ell P_R \slashed{p}]=-m_\phi^2$ and the Dirac $\delta$-functions dictate the energy threshold: $p_0 p_{\ell 0}<0$.  
The integration limit of $p_\ell\equiv |\vec p_\ell|$ comes from the angular integral via the  Dirac $\delta$-function in $G_\phi^{\gtrless}$.  The washout effect from Eq.~\eqref{eq:W-trace2} exhibits the expected scaling $f_\alpha-\bar f_{\alpha}$, which,  in the absence of the CP-violating source, implies  that   $\Delta n_\alpha$ will   be diluted exponentially.

In the weak washout regime where $\nu_{R\alpha}$  is not initially present in the thermal plasma, both $f_\alpha$ and $\bar f_\alpha$ are negligible due to the small Yukawa couplings. In the early stage, there is no  significant generation of CP asymmetries.  When $T\lesssim m_\phi$, $f_\alpha$, $\bar f_\alpha$ and their difference   increase.  In the limit of  $m_\phi \gg T$,  the $p_\ell$-integration would yield
\begin{align}
	\int_{m_\phi^2/(4p)}^\infty \left[f^{\rm eq}_\phi(p_\ell+p)+f^{\rm eq}_\ell(p_\ell)\right]=T \ln \left(\frac{e^{m_\phi^2/(4p T)}+1}{e^{m_\phi^2/(4pT)+p/T}-1}\right)+p\approx 0\,,
\end{align}
where  $p=m_\phi/2$ was used in the last approximation.  
Therefore, we see that in the weak washout regime $\mathcal{W}$ is suppressed by $f_\alpha-\bar f_\alpha$ at early times and then  it becomes suppressed by the  $p_\ell$-integration at later times even if $f_\alpha$ and $\bar f_\alpha$ can approach the equilibrium state.  

 However, if right-handed neutrinos reach thermal equilibrium already at $T>m_\phi$, the washout effect would become strong and the depletion of the generated $\nu_R$ asymmetries towards the end of neutrinogenesis should be taken into account. 
 Once right-handed neutrinos become fully thermalized, the CP-violating source would vanish, leaving the washout rate $\mathcal{W}$ to exponentially dilute the $\nu_R$ asymmetry. To see this, we 
 notice
 \begin{align}\label{eq:f-n}
 f_\alpha(p)-\bar f_\alpha (p)\approx 12 \frac{n_\alpha-\bar n_\alpha}{T^3} \frac{e^{p/T}}{(e^{p/T}+1)^2}\,,
 \end{align}
  at leading order of a  small chemical potential.  Replacing $d/dt$ with $\partial /\partial t+3\mathcal{H}$ in Eq.~\eqref{eq:KB}, where $\mathcal{H}$ is  the Hubble parameter  
  \begin{align}\label{eq:Hubble}
  	\mathcal{H}\approx 1.66\sqrt{g_\rho(T)}\frac{T^2}{m_{\rm P}}\,,
  \end{align}
  with $m_{\rm P}\approx 1.22\times 10^{19}$~GeV the Planck mass and $g_\rho(T)$ the effective degrees of freedom in energy,\footnote{We will take $g_\rho\approx g_s$ in the computation of $Y_\nu$. } we can 
evaluate   the washout  with   $\mathcal{S}_{\rm CP}=0$,  giving rise to 
  \begin{align}\label{eq:Yf-Yi}
  	Y_{\nu_\alpha}(z_f)&=Y_{\nu_\alpha}(z_i) e^{-\tilde{\mathcal{W}} }\,,\qquad
  	\tilde{\mathcal{W}}=\left(\frac{|y_\alpha|}{10^{-6}}\right)^2\left(\frac{8.45~\text{TeV}}{m_\phi}\right)I(z_i, z_f)\,,
  	\\[0.3cm]
  	I(z_i, z_f)&=\int_{z_i}^{z_f}z^2dz \int_0^\infty dx_\alpha \int_{z^2/(4x_\alpha)}^\infty dx_\ell  \frac{e^{x_\alpha}}{(e^{x_\alpha}+1)^2} \left[f^{\rm eq}_\phi(x_\ell+x_\alpha)+f^{\rm eq}_\ell(x_\ell)\right].\label{eq:If}
  \end{align}
 $Y_{\nu_\alpha}(z_i)$ is  the   $\nu_{R\alpha}$ asymmetry initially generated  at $z_i\equiv m_\phi/T_i$, and  $Y_{\nu_\alpha}(z_f)$ is the final asymmetry  at $z_f\equiv m_\phi/T_{\rm sph}$. Here we introduce the dimensionless variables    $x_\alpha \equiv p/T$ and $x_\ell\equiv p_\ell/T$. 

\begin{figure}[t]
	\centering
	\includegraphics[scale=0.55]{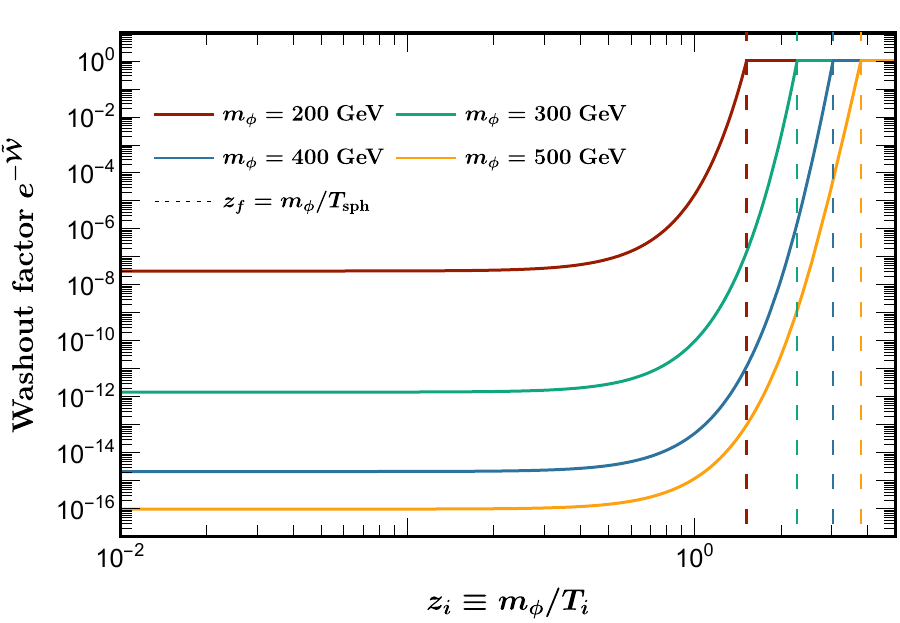} 
	\caption{\label{fig:washout_2} The   washout factor $e^{-\tilde{\mathcal{W}}}$ towards sphaleron decoupling, where forbidden neutrinogenesis completes at $z_i=m_\phi/T_i$.  The vertical lines   correspond to      sphaleron decoupling at $z_f=m_\phi/T_{\rm sph}$. }
\end{figure}

We show in Fig.~\ref{fig:washout_2} the washout effect in terms of $z_i$ and $m_\phi$ by fixing $|y_\alpha|=10^{-6}$. We see that when neutrinogenesis completes at $z_i<1$ due to $\nu_R$ thermalization, a lighter scalar generally results in a smaller dilution factor  $e^{-\tilde{\mathcal{W}}}$ and hence a smaller washout effect.  However, a scalar as light as 200~GeV would  still predict a dilution of $\nu_R$ asymmetry by a factor of  $10^3$ if $\nu_R$ has already reached thermal equilibrium at $z_i<1.22$. To ensure the washout effect with a dilution factor less than, \textit{e.g.}, $10$, $z_i\gtrsim 1.42$ is required for $m_{\phi}=200$~GeV, which is  close to sphaleron decoupling at $z_f=1.52$. When the scalar mass increases to 500~GeV, a huge dilution factor $10^{16}$ would arise if neutrinogenesis completes at $z_i<1$, rendering a vanishing $\nu_R$ asymmetry. A  dilution factor less than $10$ for $m_\phi=500$~GeV requires neutrinogenesis completes after $z_i\approx 3.61$, which however cannot be realized due to earlier  $\nu_R$ thermalization, as will be discussed in section~\ref{sec:fnu_solver}. While this could be resolved by taking a smaller Yukawa coupling,  the CP-violating source will  also  be suppressed   at the same time.

The results shown in Fig.~\ref{fig:washout_2} will provide an easy way to check whether the strong washout effect arises for a given scalar mass. This circumvents the numerical computation of the full integro-differential KB equation.

\subsection{Two-loop CP-violating source from the  PMNS phase}\label{sec:CPsource}
In this section, we will first present the  general formula for the CP-violating source. Then we discuss the  dependence of CP asymmetries on the Yukawa phase, which is unique in the minimal neutrinophilic scalar scenario.   After that, we will elaborate the general result   in section~\ref{sec:nu_asym}, after the nonthermal conditions are specified in section~\ref{sec:fnu_solver}.

\begin{figure}[t]
	\centering
	\includegraphics[scale=0.8]{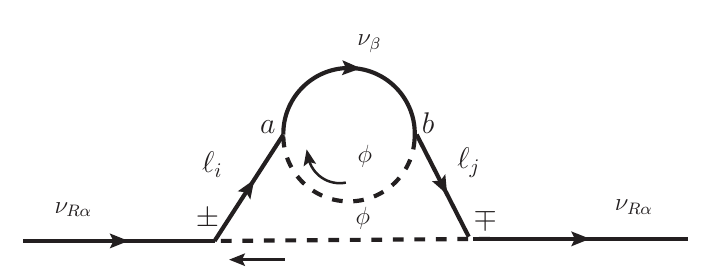} 
	\caption{\label{fig:2loop_CP} The two-loop self-energy diagram of $\nu_{R\alpha}$ contributing to the forbidden CP asymmetry.  The outer thermal indices $\pm$ correspond to the amplitudes $\slashed{\Sigma}_{\nu_\alpha}^{\lessgtr}$. The inner vertices $a,b$ are summed over thermal indices $\pm$, generating terms given in Eqs.~\eqref{eq:tildeS<Lij}-\eqref{eq:tildeS>Lij}. We use the arrows for the scalar propagators to denote the momentum flow, and  make  the fermion arrows align with the momentum flow.  }
\end{figure}

In the SK-CTP formalism, the leading-order CP-violating source starts at two-loop level.  
The two-loop   self-energy diagrams that can induce $\nu_R$ asymmetries  are determined by Fig.~\ref{fig:2loop_CP}. The amplitudes read
\begin{align}\label{eq:S<_chi1}
	i\slashed{\Sigma}_{\nu_\alpha}^<(p_{})&=2y_4\int_{p_\ell}\int_{p_\phi} (2\pi)^4\delta^4(p_{}-p_\ell+p_\phi)P_L i\slashed{{S}}^<_{\ell_{ij}}P_R iG_\phi^>\,,
	\\[0.3cm]
	i\slashed{\Sigma}_{\nu_\alpha}^>(p_{})&=2y_4^* \int_{p_\ell}\int_{p_\phi} (2\pi)^4\delta^4(p_{}-p_\ell+p_\phi)P_L i\slashed{S}^>_{\ell_{ji}}P_R iG_\phi^<\,,\label{eq:S>_chi1}
\end{align}
where the factor of 2 comes from gauge $SU(2)_L$ degeneracy. The Yukawa coupling $y_4$ is defined by
\begin{align}\label{eq:Y4}
	y_4\equiv y_{i\alpha}y^{*}_{j\alpha}y_{i\beta}^{*}y_{j\beta}\,,
\end{align}
and we have played the trick of interchanging   dummy indices $i, j$ in $i\slashed{\Sigma}_{\nu_\alpha}^>$ such that the dependence on the Yukawa couplings is complex-conjugated to that in $i\slashed{\Sigma}_{\nu_\alpha}^<$. Such a difference will lead to the common expectation that the CP-violating source   depends on  the Yukawa phase, i.e., $\text{Im}(y_4)$.  

In Eqs.~\eqref{eq:S<_chi1}-\eqref{eq:S>_chi1}, $i\slashed{S}^{<}_{\ell_{ij}}$ and $i\slashed{S}^{>}_{\ell_{ji}}$ comprise  the  product of  resummed lepton propagators and the inner loop amplitudes.  Summing  the thermal indices $a,b=\pm$ in the inner loop gives arise to\footnote{See also the Supplemental Material in Ref.~\cite{Kanemura:2024dqv}.}
	\begin{align}\label{eq:tildeS<Lij}
	i\slashed{{S}}^<_{\ell_{ij}}&=(i\slashed{{S}}^R_{\ell_i}) (-i\slashed{\Sigma}^T_{\ell_{}})(i\slashed{{S}}^<_{\ell_j})+(i\slashed{{S}}^R_{\ell_i})(-i\slashed{\Sigma}^<_{\ell_{}})(i\slashed{{S}}^R_{\ell_j})+(i\slashed{{S}}^<_{\ell_i})(-i\slashed{\Sigma}^<_{\ell_{}})(i\slashed{{S}}^R_{\ell_j})
	\nonumber\\[0.2cm]
	&-(i\slashed{{S}}^R_{\ell_i})(-i\slashed{\Sigma}^<_{\ell_{}})(i\slashed{{S}}^>_{\ell_j})-(i\slashed{{S}}^<_{\ell_i})(-i\slashed{\Sigma}^{\bar T}_{\ell_{}})(i\slashed{{S}}^R_{\ell_j})\,,
	\\[0.2cm]
	i\slashed{{S}}^>_{\ell_{ji}}&=(i\slashed{{S}}^R_{\ell_i})(-i\slashed{\Sigma}^T_{\ell})(i\slashed{{S}}^>_{\ell_j})+(i\slashed{{S}}^R_{\ell_i})(-i\slashed{\Sigma}^>_{\ell})(i\slashed{{S}}^R_{\ell_j})+(i\slashed{{S}}^<_{\ell_i})(-i\slashed{\Sigma}^>_{\ell})(i\slashed{{S}}^R_{\ell_j})
	\nonumber\\[0.2cm]
	&-(i\slashed{{S}}^R_{\ell_i})(-i\slashed{\Sigma}^>_{\ell})(i\slashed{{S}}^>_{\ell_j})-(i\slashed{{S}}^>_{\ell_i})(-i\slashed{\Sigma}^{\bar T}_{\ell})(i\slashed{{S}}^R_{\ell_j})\,,\label{eq:tildeS>Lij}
\end{align}  
where  $\slashed{{S}}^{R, \lessgtr}_{\ell_i}$ denote the resummed retarded and Wightman propagators of lepton $i$,  and   the Yukawa couplings from the inner loop amplitudes  $-i\slashed{\Sigma}^{ab}_{\ell_{ij}}$ have been factored out and absorbed into $y_4$, with
\begin{align}
\slashed{\Sigma}^{ab}_{\ell_{ij}}\equiv y_{i\beta}^* y_{j\beta} \slashed{\Sigma}^{ab}_{\ell}\,, \quad  a,b =\pm\,.
\end{align}

To calculate the CP asymmetry with resummed thermal leptons, it is beneficial  to take a proper basis where   thermal corrections to  leptons are diagonal in flavor space. To this aim, we choose the basis where the charged-lepton Yukawa matrix is diagonal. When the thermal mass  correction to leptons  is dominated by the SM contributions (gauge and charged-lepton Yukawa interactions),  the basis  we choose will   make both the lepton thermal mass matrix and   the resummed lepton propagators    diagonal in   flavor space~\cite{Li:2021tlv}. In this basis,  the neutrino Yukawa matrix $y$ given in Eq.~\eqref{eq:lag} and the physical PMNS matrix would be related via\footnote{Recall that the PMNS matrix is defined in the weak charged current: $ \bar e_i \gamma^\mu P_L U_{ij}\nu_{j} W_\mu^++\rm h.c.$ for $e_i=e, \mu, \tau$. }
\begin{align}
	y_{ij}=\frac{\sqrt{2}}{v_2}U_{ij}m_j \,,
\end{align}
where $m_j$ represents  the   physical neutrino masses, and we have used unitary PMNS matrix $U^\dagger U=U U^\dagger =1$. In 
the standard parameterization, $U$ is given by~\cite{ParticleDataGroup:2024cfk}
\begin{align}
	U=\left(
	\begin{array}{ccc}
		c_{{12}}c_{{13}} \,  &  s_{{12}} c_{{13}} \,   &   e^{-i \delta_{\rm CP}}   s_{{13}} \\[0.15cm]
		-s_{{12}}c_{{23}}-e^{i \delta_{\rm CP}}c_{{12}} s_{{13}} s_{_{23}}   \, &    	c_{{12}}c_{{23}}-e^{i \delta_{\rm CP}}s_{{12}} s_{{13}} s_{{23}}\,&   c_{{13}} s_{{23}} \\[0.15cm]
		s_{{12}}s_{{23}}-e^{i \delta_{\rm CP}}c_{{12}} s_{{13}} c_{{23}}  \, & 	-c_{{12}}s_{{23}}-e^{i \delta_{\rm CP}}s_{{12}} s_{{13}} c_{{23}}\,  & c_{{13}}c_{{23}}\\
	\end{array}
	\right),
\end{align}
where $s_{ij}\equiv \sin\theta_{ij}$ and $c_{ij}\equiv \cos\theta_{ij}$ correspond to the mixing angles, while  $\delta_{\rm CP}$ is the Dirac CP-violating phase that could feature large CP violation in the lepton sector. 
Then we can rewrite the Yukawa couplings as 
\begin{align}\label{eq:yalpha_sq}
	|y_{\alpha}|^2&\equiv \sum_i y_{i\alpha}y^*_{i\alpha}=\frac{2}{v_2^2}|U_{i\alpha}|^2 m_\alpha^2\,,
	\\[0.2cm]
	\text{Im}(y_4)&=\frac{4}{v_2^4}\text{Im}\left(U_{i\alpha}U^*_{j\alpha}U^*_{i\beta}U_{j\beta}\right)m_\alpha^2 m_\beta^2\,,\label{eq:Imy4}
\end{align}
where $\text{Im}(y_4)$ represents the Yukawa phase for  the CP asymmetry. Clearly, the Dirac CP-violating phase provides the unique source for CP violation in generating the $\nu_R$ asymmetry.   If   CP conservation from neutrino oscillation experiments is confirmed, forbidden neutrinogenesis will not work to explain the BAU in the minimal neutrinophilic scalar scenario. 

Notice   that  in Eq.~\eqref{eq:Imy4} the   indices $i,j, \alpha, \beta$ should not  be summed trivially, since the CP-violating source will carry additional dependence on these indices from lepton thermal masses and neutrino distribution functions, both of which are crucial to induce a nonzero  CP-violating source. 
In fact, if the indices $i,j$  can be summed trivially in Eq.~\eqref{eq:Imy4}, we would arrive at    $\sum_{i,j}\text{Im}(y_4)=0$. However, the flavor-dependent lepton thermal masses introduce additional $i,j$ dependence, which,   arising  from Eqs.~\eqref{eq:tildeS<Lij}-\eqref{eq:tildeS>Lij},  is a key point in forbidden neutrinogenesis, and more generally in lepton-number conserving forbidden leptogenesis.  We should   emphasize  that the   lepton thermal masses are not inserted by hand,\footnote{This treatment is often  taken in the calculation of $S$-matrix amplitudes, where thermal masses are empirically inserted into the Feynman propagators~\cite{Li:2021tlv}. } but    is a consistent requirement   from the  two-particle-irreducible effective action constructed in the SK-CTP formalism~\cite{Berges:2004yj,Kanemura:2024dqv}.

\subsection{Nonthermal conditions}\label{sec:fnu_solver}

In addition to the PMNS phase  as the source for CP violation,  the nonthermal condition is also mandatory to realize leptogenesis in the early universe. Before elaborating  the CP-violating source, we should determine the nonthermal conditions in  the minimal neutrinophilic scalar scenario.

 First of all, unlike the SM Higgs boson, the  neutrinophilic scalar has a vacuum mass ($m_\phi=\mu_2$) before   electroweak gauge symmetry breaking.  Under cosmic expansion, the mass   will drive the scalar into the out-of-equilibrium  regime   when the temperature drops to $m_\phi$, providing a nonthermal condition for leptogenesis.
  However, to ensure an asymmetry in the  $\nu_R$ sector, at least one of the right-handed neutrinos must  be also out of equilibrium~\cite{Kanemura:2024dqv}.   Then how many $\nu_R$ flavors are sufficient to  generate the requisite BAU? To answer this,  we should recall the neutrino mass spectrum observed in neutrino oscillations~\cite{Esteban:2024eli}. Currently, the lightest neutrino mass, either in   normal ordering or inverted ordering, is not   determined yet by experiments, but it has an upper limit when imposing the  bound  from cosmology $\sum_i m_i<0.12$~eV~\cite{Planck:2018vyg}, which is $m_{1,\rm max}\approx 0.03$~eV in the normal ordering and $m_{3,\rm max}\approx 0.02$~eV in the inverted ordering. For the lightest neutrino mass ranging from the maximal value to zero, the other two heavier neutrinos do not exhibit strong mass hierarchy. Given Eq.~\eqref{eq:nu-mass}, the Yukawa couplings of the two heavier neutrinos would be at the similar order,  indicating that they would basically  follow the same evolution in the early universe. Therefore, in addition to a nonthermal scalar, we can either  have the lightest $\nu_R$ or all the three $\nu_R$  be nonthermal during neutrinogenesis.   
  
  Nevertheless, three $\nu_R$ being out of equilibrium implies that their Yukawa couplings should all be small enough, where the CP-violating source could be suppressed. In this case, increasing the Yukawa coupling should be taken carefully since the right-handed neutrinos may reach thermal equilibrium towards the end of neutrinogenesis, where the washout effect discussed in section~\ref{sec:washout} will readily dilute the $\nu_R$ asymmetries.  
  
   When the two heavier $\nu_R$  have larger Yukawa couplings such that they already reached thermal equilibrium at neutrinogenesis, there would be  a Yukawa coupling enhancement in the CP-violating source.  In addition, large Yukawa couplings will open more detection channels to test the  neutrinophilic scalar scenario.

   In either case, it is helpful to estimate the maximal neutrino  Yukawa coupling that can delay neutrino thermalization.
Since right-handed neutrinos are gauge singlets, the evolution  of $\nu_R$ is well determined  by decay and inverse decay of the  neutrinophilic scalar. In the SK-CTP formalism,  the one-loop neutrino self-energy diagrams   shown in Fig.~\ref{fig:washout} can also be used to determine the abundance of right-handed neutrinos. Given that    the two-loop  diagram shown in Fig.~\ref{fig:2loop_CP} determines the   number asymmetry $n_\nu-n_{\bar \nu}$,  we can neglect the small asymmetry $f_\alpha-\bar f_\alpha$  when calculating the time evolution of $f_\alpha$ from Fig.~\ref{fig:washout}. Under this approximation, the Boltzmann equation usually  suffices to capture the solution of $f_\alpha$, which is easier to manage since the amplitudes of decay and inverse decay  are obtained from tree-level diagrams. For consistent checks,   we will still apply  the KB equation of $f_\alpha$.  To this end, 
we first multiply $\text{sign}(p_0)$ on both sides of the KB equation given in Eq.~\eqref{eq:0th-KB},   take the  Dirac trace, and then integrate over $p_0$.\footnote{To avoid confusion, we should emphasize that this treatment is different from  Eq.~\eqref{eq:W-trace2}  and will not lead to a collision rate   scaling as $f_\alpha-\bar f_\alpha$.} We arrive at 
\begin{align}
	\frac{df_\alpha}{dt}&=\frac{1}{8\pi}\int dp_0 \text{sign}(p_0)\text{Tr}\left[i\slashed\Sigma_{\nu_\alpha}^> i\slashed S_{\nu_\alpha}^<-i\slashed\Sigma_{\nu_\alpha}^<i \slashed S_{\nu_\alpha}^>\right]_{\rm 1-loop}
\nonumber	\\[0.2cm]
	&=\frac{|y_\alpha|^2m_\phi^2}{32\pi p^2}\int_{m_\phi^2/(4p)}^\infty dp_\ell I(p_\ell, p),\label{eq:dfalpha/dt}
\end{align}
where the integration limit $p_\ell>m_\phi^2/(4p)$ reflects energy-momentum conservation $p^\mu_\phi=p^\mu_\ell+p^\mu$ in decay/inverse decay. The statistics function is given by 
\begin{align}\label{eq:statistics0}
	I(p_\ell, p)\equiv f^{\rm eq}_\phi(p_\ell+p)\left[1-f^{\rm eq}_\ell(p_\ell)\right]-f_\alpha (p)\left[f^{\rm eq}_\phi(p_\ell+p)+f^{\rm eq}_\ell(p_\ell)\right],
\end{align}
where we have taken the thermal distribution functions for the scalar and leptons. 
 Rearranging the above statistics function, we can check that it is equivalent to  that from Boltzmann equation: $f^{\rm eq}_\phi(1-f^{\rm eq}_\ell)(1-f_\alpha)-f^{\rm eq}_\ell f_\alpha(1+f^{\rm eq}_\phi)$. This confirms  that the KB equation can derive the Boltzmann equation when the KB ansatz and quasiparticle approximation are applied~\cite{Buchmuller:2000nd,Prokopec:2003pj,Prokopec:2004ic}.

\begin{figure}[t]
	\centering
	\includegraphics[scale=0.35]{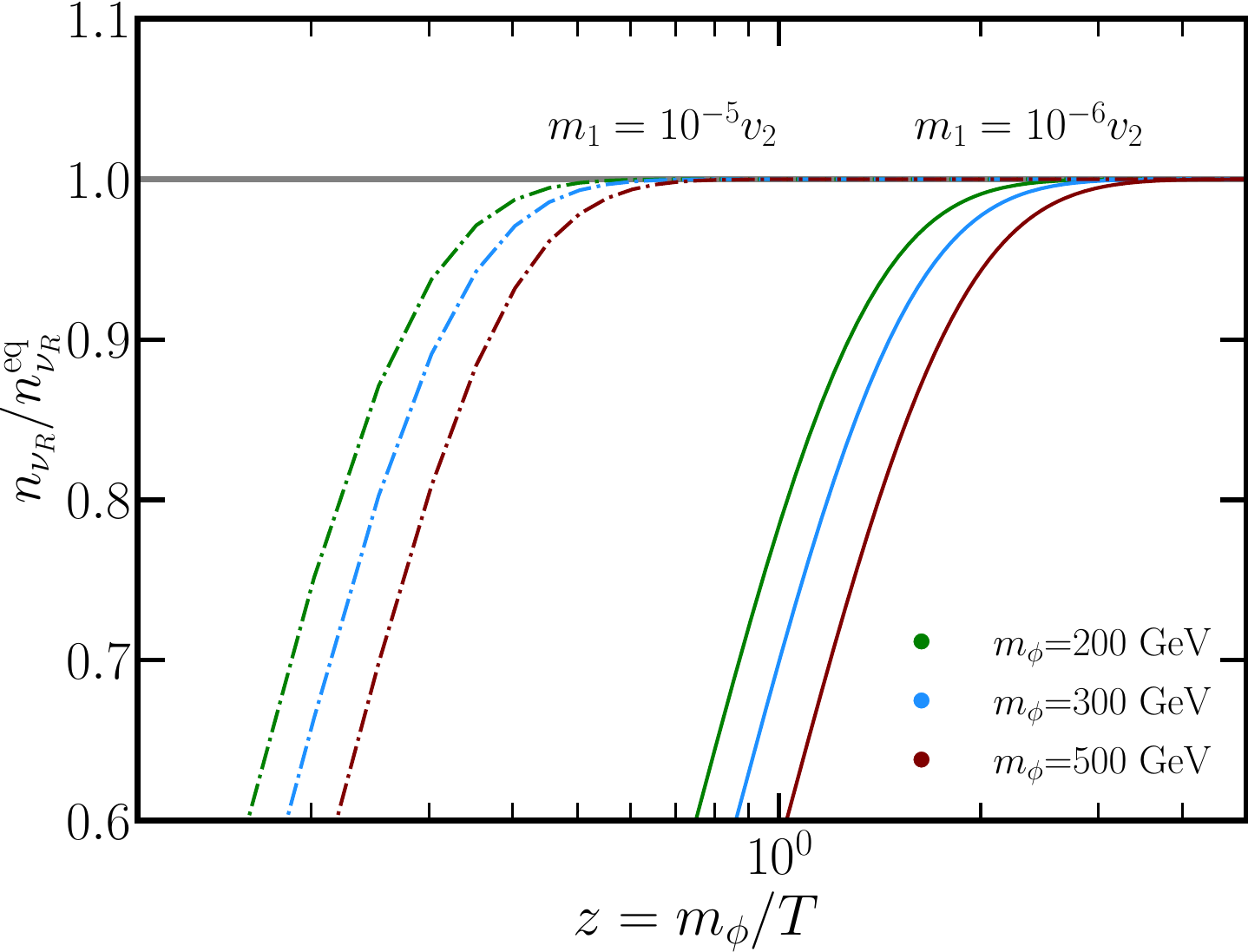} 
	\caption{\label{fig:f1} The evolution of  $\nu_{R1}$ from $\phi$ decay and inverse decay, normalized to the thermal equilibrium limit. The benchmark point $m_1/v_2=10^{-6} (10^{-5})$  typically  corresponds  to the   situation where $\nu_{R1}$  thermalization occurs after (before) $T=m_\phi$. Here,  we take the normal ordering for illustration, where $m_1$ corresponds to the lightest neutrino mass.  }
\end{figure}

 For concreteness, we take $\alpha=1$ to see the evolution of the neutrino number density normalized to the equilibrium limit:
 \begin{align}
 	n_{\nu_R}^{\rm eq}=\int\frac{d^3 p}{(2\pi)^3} \frac{1}{e^{p/T}+1}\,.
 \end{align}
We show in Fig.~\ref{fig:f1} the normalized neutrino number density in terms of the time variable $z\equiv m_\phi/T$. We choose two special values  $m_1/v_2=10^{-6}, 10^{-5}$ to illustrate the  moments when   $\nu_{R1}$ reaches full thermal equilibrium.  We see that  $m_1/v_2=10^{-6}$  typically  results in  $\nu_R$  thermalization after $T=m_\phi$ while $m_1/v_2=10^{-5}$ leads to   earlier thermalization   unless the scalar mass becomes large. Nevertheless, forbidden neutrinogenesis from  a heavier scalar implies that there would be a longer period between the completion of neutrinogenesis and   sphaleron decoupling. This would still  lead to a strong washout effect even if $\nu_R$ thermalization occurs after $T=m_\phi$. 

If forbidden neutrinogenesis completes before    sphaleron decoupling due to $\nu_R$ thermalization, we should  check whether the strong washout of $\nu_R$ asymmetries occurs.  The thermalization moments shown in Fig.~\ref{fig:f1}, together with the dilution exponent shown in Fig.~\ref{fig:washout_2}, will provide  such  checks in the next section.  In particular, avoiding the strong washout effect generally requires $m_\phi<500$~GeV for  a neutrino Yukawa coupling $|y|=10^{-6}$.

\subsection{Neutrino asymmetries}\label{sec:nu_asym}

\subsubsection{Three nonthermal neutrinos}\label{sec:Y3}
We first consider the CP-violating source in the case of three nonthermal $\nu_R$.  We will work at   leading order of $\delta f\equiv f-f^{\rm eq}$ and neglect  the quadratic correction $\delta f_\phi \delta f_{\nu_R}$ from both the nonthermal scalar and neutrinos. Therefore, for  three nonthermal $\nu_R$, we will take the thermal  distribution for the scalar.  We relegate the detailed calculations of the CP-violating rate  to   Appendix~\ref{append:SCP_EWscalar}.  The final result   is rather simple and reads
\begin{align}\label{eq:S_CP_fin}
	\mathcal{S}_{\rm CP}=\frac{\text{Im}(y_4)m_\phi^4}{256\pi^4 (\tilde{m}_j^2-\tilde{m}_i^2)}&\int^\infty_0 \frac{dp_\ell}{p_\ell}\int^\infty_{\frac{m_\phi^2}{4p_\ell}+p_\ell} dE_\phi  \int^\infty_{\frac{m_\phi^2}{4p_\ell}} dq I(p_\ell, E_\phi, q)\,,
\end{align}
where the statistics function gives
\begin{align}\label{eq:statistics}
	I(p_\ell, E_\phi, q)=\delta f_\beta(q) [f^{\rm eq}_\phi(E_\phi)-f_\alpha(E_\phi-p_\ell)] [f^{\rm eq}_{\phi}(q+p_\ell)+f^{\rm eq}_\ell(p_\ell)]\,.
\end{align}
Given Eqs.~\eqref{eq:Y4} and~\eqref{eq:Imy4}, we see that if the distribution functions for right-handed neutrinos are flavor universal, $\mathcal{S}_{\rm CP}$  would vanish after summing over neutrino flavors $\alpha,\beta$, and hence the total CP asymmetry in the  $\nu_R$ sector vanishes. However, it is generally not the case due to the flavor-dependent Yukawa couplings.  

The statistics function given in Eq.~\eqref{eq:statistics} clearly dictates that no CP-violating source exists if all the particles are in thermal equilibrium. On the other hand, since the inner-loop scalar is heavier than the external leptons,   the inner loop particles cannot go on-shell in vacuum due to the kinetic threshold. This is understood by the optical theorem in vacuum quantum field theory that the external leptons   decay to the heavier scalar is kinematically forbidden. However, the  thermal distributions of particles in the plasma open more energy-momentum conserved emission and absorption processes~\cite{Weldon:1983jn}, with the transition probability weighted by these distributions.  This can be simply seen    if we drop the  distribution functions of right-handed neutrinos in the inner loop, which amounts to neglecting the finite-density background of right-handed neutrinos, the CP-violating source $\mathcal{S}_{\rm CP}$ would vanish. This dependence on distribution functions from inner-loop particles  demonstrates the very nature of forbidden leptogenesis.

From the result of   $I(p_\ell, E_\phi, q)$, it is worth mentioning that the dependence of the distribution functions from the outer loop  is linear in $f_\phi-f_\alpha$   and  from the inner loop is linear in $f_\phi+f_\ell$. Such linear dependence is a consequence of calculating the two-loop amplitudes under the KB equation.  It was previously    found to be  different from the quadratic dependence obtained by applying  the finite-temperature time-ordered cutting rules to $S$-matrix amplitudes~\cite{Giudice:2003jh} unless the   retarded/advanced cutting rules are  properly  used~\cite{Garny:2010nj,Frossard:2012pc,Li:2020ner}. However,  the calculation of  forbidden leptogenesis via the $S$-matrix formalism embedded in the semi-classical Boltzmann equations can also lead to inconsistent conclusions. In section~\ref{sec:KBvsBol}, we will make a comparison   between the calculations of two-loop self-energy diagrams under the KB equation and of one-loop self-energy diagrams under the Boltzmann equation, pointing out that forbidden leptogenesis within the Boltzmann approach   also  suffers from the real-intermediate-subtraction issue~\cite{Kolb:1979qa,Giudice:2003jh}.

In the weak washout regime, we may neglect the washout rate such that the particle-number asymmetry stored in  three right-handed Dirac neutrinos would  simply read 
\begin{align}\label{eq:Ynu}
	Y_{\nu}=\sum_\alpha Y_{\nu_{R\alpha}} =\int_{T_{\rm sph}}^\infty \frac{\mathcal{S}_{\rm CP}}{s_{\rm SM} \mathcal{H} T} dT\,.
\end{align}
where the entropy density and the  Hubble parameter are given by Eq.~\eqref{eq:entropy} and Eq.~\eqref{eq:Hubble}, respectively, and we have cut the temperature integration at the sphaleron decoupling moment $T_{\rm sph}\approx 132$~GeV.\footnote{The modification to the SM prediction of $T_{\rm sph}$ depends on the $\lambda$ parameters given in the Higgs potential~\eqref{eq:pot}, which we assume to be small here for simplicity. We mention that a larger $T_{\rm sph}$ will reduce $Y_\nu$, but will also suppress the washout effect. }  The right-hand side of Eq.~\eqref{eq:Ynu} is a 4-dimensional integral, where the Monte Carlo algorithm is sufficient to perform the numerical integration.

In Eq.~\eqref{eq:Ynu}, the dependence of $Y_\nu$  on the PMNS phase and the neutrino mass spectrum opens  an avenue to test the minimal forbidden neutrinogenesis.
Measurements of PMNS elements from neutrino oscillation in recent years have suggested some correlation between the sign of $\sin\delta_{\rm CP}$ and the neutrino mass ordering~\cite{Esteban:2024eli}. In the normal ordering (NO), where $m_1<m_2<m_3$, the maximal CP violation with $\delta_{\rm CP}=\pi/2$ is disfavored at $3\sigma$ level. In the inverted ordering (IO), where $m_3<m_1<m_2$, the  maximal CP violation with  $\delta_{\rm CP}=3\pi/2$ is favored    by both the T2K~\cite{T2K:2023smv,T2K:2024wfn} and NO$\nu$A~\cite{NOvA:2023iam} experiments. In addition,   $0<\delta_{\rm CP}<\pi$ is disfavored in the IO pattern within the $3\sigma$ range of data.  Based on Eq.~\eqref{eq:S_CP_fin}, we  will first  analyze  the dependence of the CP-violating source on   $\delta_{\rm CP}$ and the lightest neutrino mass, and  then see how measurements of these observables   can help to probe    forbidden neutrinogenesis in the minimal  scenario. 

We take the data from NuFIT~6.0~\cite{Esteban:2024eli}. We  fix the mixing angles with their best-fit points, which do not deviate significantly between the NO and the IO, and take  the two squared mass differences with their central values. We apply the upper bound on the sum of neutrino masses from Planck~\cite{Planck:2018vyg}, with $\sum_i m_i<0.12$~eV. For each ordering pattern, we take the lightest neutrino mass with a value    approaching  the upper bound and a much smaller one.  We further estimate $\mathcal{S}_{\rm CP}$ by taking the maximal resonant enhancement from the muon Yukawa coupling.\footnote{This corresponds to $j=2,i=1$ in Eq.~\eqref{eq:S_CP_fin}. Note that the result from $j=1,i=2$  is identical to  $j=2,i=1$ since $\text{Im}[y_4(j=1,i=2)]=-\text{Im}[y_4(j=2,i=1)]$. } Let us define
\begin{align}\label{eq:S_CP_est}
	\mathcal{S}_{\rm CP}\equiv \frac{\text{Im}(y_4)}{8\pi^4 y_\mu^2}\langle f_\beta\rangle\,,
\end{align}
where $\langle \delta f_\beta\rangle$ is a dimensionful  average of $\delta f_\beta$ under the integration of Eq.~\eqref{eq:S_CP_fin}. $\langle \delta f_\beta\rangle$    depends sensitively on the neutrino flavor $\beta$, but weakly   on flavor $\alpha$. Given this, we neglect $f_\alpha$ from Eq.~\eqref{eq:statistics} at the moment  to estimate Eq.~\eqref{eq:S_CP_est}, but will include  $f_\alpha$ at  a later stage.

\begin{table}
	\centering
	\renewcommand{\arraystretch}{1.4}
	\begin{tabular}{l|c|c}
		\hline\hline
		Patterns &  ($m_1,m_2, m_3$) (meV) & $10^{16}\mathcal{S}_{\rm CP}/\sin\delta_{\rm CP}$~($\rm eV^4$) \\ 
		\hline 
		NO+$m_{1,\rm max}$        &  	 (30.14\,,\, 31.36\,,\, 58.49)   &  
		$1.21\left(\langle \delta f_3\rangle-\langle \delta f_2\rangle\right)$ \\
		\hline
		NO+$10^{-3}m_{1,\rm max}$           &  	 (0.03\,,\, 8.65\,,\, 50.13)    &  $ 0.89(\langle \delta f_3\rangle-\langle \delta f_2\rangle)$  \\
		\hline
		IO+$m_{3,\rm max}$        &  	(51.63\,,\, 52.35\,,\, 16.02)   &  
		$-0.09 \left(\langle \delta f_2\rangle-\langle \delta f_3\rangle\right)$ \\
		\hline
		IO+$10^{-3}m_{3,\rm max}$       &  	  (49.08\,,\, 49.84\,,\,  0.016)  &  $-9\times 10^{-8 }(\langle \delta f_2\rangle- \langle\delta f_3\rangle) $  \\
		\hline	\hline
	\end{tabular}
	\caption{Behavior of $\mathcal{S}_{\rm CP}$ in terms of the  neutrino mass spectrum and Dirac CP-violating phase $\delta_{\rm CP}$ in the case of three nonthermal $\nu_R$. Pattern NO+$m_{1,\rm max}$ (IO+$m_{3,\rm max}$)  denotes the normal (inverted) ordering with a maximally allowed mass for the lightest neutrino,  and NO+$10^{-3}m_{1,\rm max}$ (IO+$10^{-3}m_{3,\rm max}$)   denotes the normal (inverted) ordering but with a   smaller value for the lightest neutrino mass.  The mass spectrum in NO (IO) gives $m_1<m_2<m_3$ ($m_3<m_1<m_2$), where   $\delta f_3$ corresponds to the heaviest (lightest) neutrino. $\mathcal{S}_{\rm CP}$  is estimated via Eq.~\eqref{eq:S_CP_est}, and  $v_2=1$~keV is chosen, which  only affects the overall magnitude of  $\mathcal{S}_{\rm CP}$.  }
	\label{tab:neu-info}
\end{table}

Following the above setup, we show in Tab.~\ref{tab:neu-info} the neutrino mass spectrum and $\mathcal{S}_{\rm CP}$ in four different patterns. In the pattern of NO+$m_{1,\rm max}$,  the first two generations have similar masses and hence we expect $\langle \delta f_1\rangle \approx \langle \delta f_2\rangle$. However, the third generation has a larger mass, indicating a larger Yukawa coupling and hence   $\langle \delta f_2\rangle<\langle \delta f_3\rangle<0$. Therefore we see that $0<\delta_{\rm CP}<\pi$ will lead to a positive $Y_B$. 

In pattern $\text{NO}+10^{-3}m_{1,\rm max}$, the lightest neutrino mass $m_1$ is taken by $10^{-3}\times m_{1,\rm max}$, which implies a much smaller Yukawa coupling for the lightest generation and hence a negligible contribution to $\mathcal{S}_{\rm CP}$.  In this case, however, the dominant contribution from the two heavier $\nu_R$ does not change significantly, so we still obtain a   result similar to NO+$m_{1,\rm max}$, where
  $0<\delta_{\rm CP}<\pi$ is responsible for a  positive $Y_B$. 

In  pattern    IO+$m_{3,\rm max}$, the heavier two generations have similar masses and hence we expect $\langle \delta f_1\rangle \approx \langle \delta f_2\rangle$. Furthermore, due to a smaller mass for the lightest neutrino, we expect $\langle \delta f_3\rangle<\langle \delta f_2\rangle<0$. It indicates that 
  $\pi<\delta_{\rm CP}<2\pi$ will lead to a positive $Y_B$. While  $\mathcal{S}_{\rm CP}$  in IO+$m_{3,\rm max}$  has a   prefactor smaller than from  NO+$m_{1,\rm max}$ by  one order of magnitude, CP violation in the IO pattern could still be maximal with $\sin\delta_{\rm CP}=-1$ and larger than in   the NO pattern with $\sin\delta_{\rm CP}= \mathcal{O}(0.1)$.
Therefore, $\mathcal{S}_{\rm CP}$ from both NO+$m_{1,\rm max}$ and IO+$m_{3,\rm max}$ could be at the same  order of magnitude. The crucial difference in these patterns is  that the condition  $Y_B>0$ requires different   signs of $\sin\delta_{\rm CP}$, where     $0<\delta_{\rm CP}<\pi$  in the NO  and $\pi<\delta_{\rm CP}<2\pi$  in the IO will be selected. This is actually in line with the implication from current measurements. Therefore, both NO and IO with the lightest neutrino mass approaching its maximal value can predict a positive baryon asymmetry. However, such degeneracy would   break down if the NO pattern with $\pi\leqslant\delta_{\rm CP}\leqslant2\pi$ is detected     in   upcoming measurements.\footnote{Actually, this pattern has already been hinted by the best-fit point of $\delta_{\rm CP}$ from NuFIT (\url{http://www.nu-fit.org}). However the favored values of $\delta_{\rm CP}$ in the NO by the T2K~\cite{T2K:2023smv,T2K:2024wfn} and NO$\nu$A~\cite{NOvA:2023iam} experiments do not agree.}

\begin{figure}[t]
	\centering
	\includegraphics[scale=0.35]{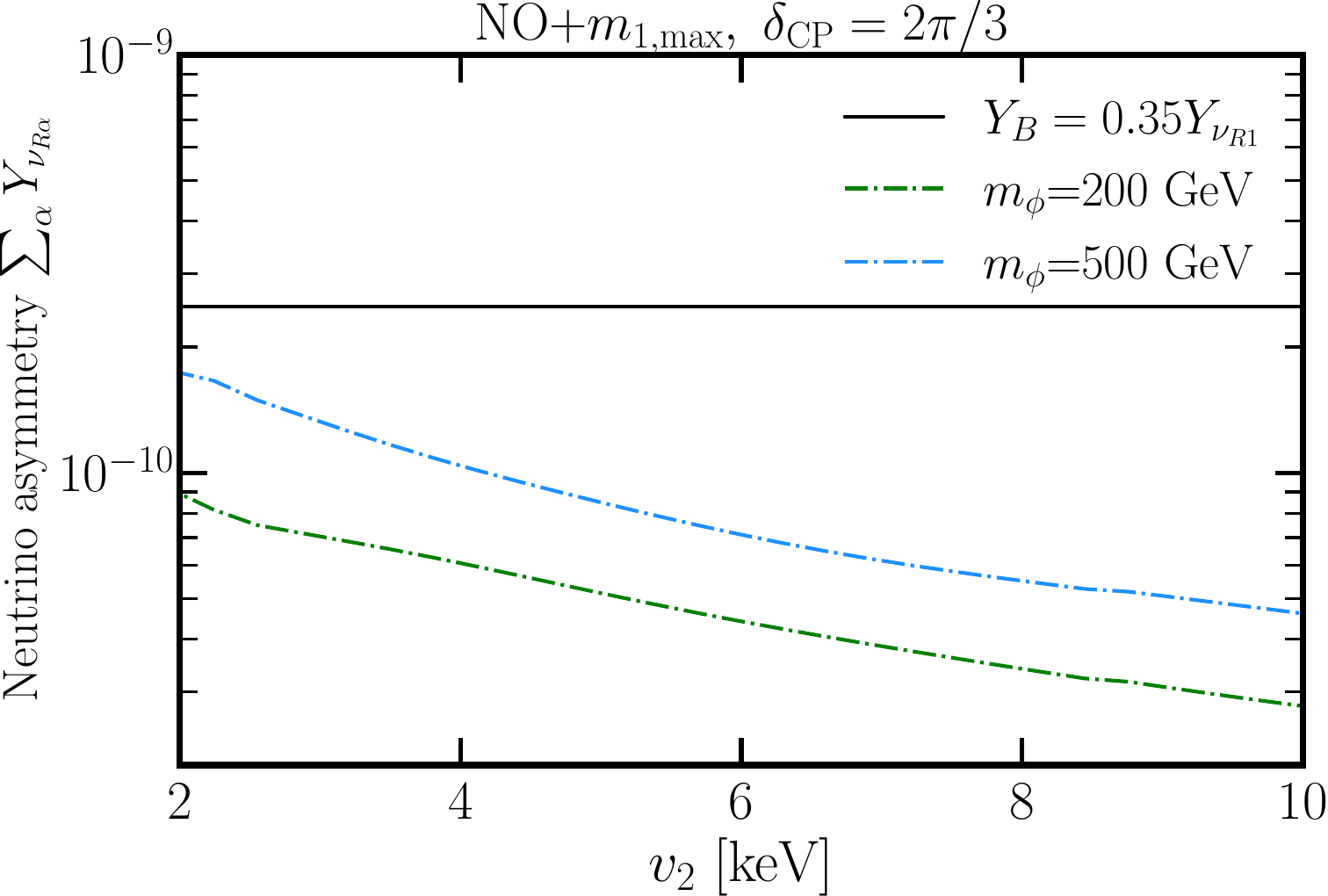}
	\caption{\label{fig:Y3} The total asymmetry in the case of three nonthermal $\nu_R$. Pattern NO with a maximal value for the lightest neutrino mass $m_{1,\rm max}\approx 0.03$~eV is taken for illustration, where the Dirac CP-violating phase is fixed by $\delta_{\rm CP}=2\pi/3$.  }
\end{figure}

In pattern $\text{IO}+10^{-3}m_{3,\rm max}$, we expect $\langle \delta f_1\rangle\approx \langle \delta f_2\rangle$ since the two heavier  generations become quasi-degenerate in the limit of vanishing $m_3$. Unlike pattern $\text{NO}+10^{-3}m_{1,\rm max}$, this quasi-degenerate mass leads to a much smaller $\mathcal{S}_{\rm CP}$, 
which will essentially vanish in the limit of $m_3=0$. 

Before the full numerical computation of Eq.~\eqref{eq:Ynu}, let us get  a rough estimate of the parameter space for $Y_{\nu}\simeq 10^{-10}$. Following Eq.~\eqref{eq:S_CP_fin}, we can generally write
\begin{align}\label{eq:Ynu_est}
	Y_\nu \sim   \text{Im}(y_4) \left( \frac{m_{\rm P}}{m_\phi}\right)\int z^4 I(x_\ell, x_\phi, x_\alpha) dx_\ell dx_\phi dx_\alpha dz\,,
\end{align}  
up to $\mathcal{O}(1)$ corrections. The dimensionless variables are defined as  $ x_\ell\equiv p_\ell/T, x_\phi\equiv p_\phi/T, x_\alpha \equiv p/T, z\equiv m_\phi/T$. Note that we have taken the maximal enhancement from the muon Yukawa coupling $y_\mu\approx 6\times 10^{-4}$, and  the 4-dimensional  integration may reach $\mathcal{O}(1)$ for large $\delta f_\beta\sim 0.1$.\footnote{Typically, the integration over Bose-Einstein and Fermi-Dirac distributions can reach $\mathcal{O}(1)$, while the integration over $z$ is dominated at $z=\mathcal{O}(1)$, corresponding to the epoch when neutrinogenesis culminates. }  Taking this $\mathcal{O}(1)$ estimate and assuming an electroweak scalar, we find that 
the order of magnitude for the neutrino Yukawa couplings should be larger than $10^{-7}$ to match $Y_{\nu}\simeq 10^{-10}$. Given Eq.~\eqref{eq:Imy4}, we have $\text{Im}(y_4)\sim U^4 m^4/v_2^4$, where  $U^4$ denotes  the    estimate of the PMNS matrix elements at the quartic power, with $U^4\sim 0.01$. It turns out that we need $m/v_2\gtrsim 10^{-5}$ to produce  $Y_{\nu}\simeq 10^{-10}$. However,  Fig.~\ref{fig:f1} suggests that $m/v_2\gtrsim 10^{-5}$ would lead to a large washout effect. 

The above estimate is confirmed by the full numerical computation of Eq.~\eqref{eq:Ynu}, as shown in  Fig.~\ref{fig:Y3}. In pattern  NO+$m_{1,\rm max}$, we see that generating the observed baryon asymmetry $Y_B$ requires $v_2<2$~keV, which corresponds to $m_1/v_2>10^{-5}$. This will introduce a large washout factor  since $\nu_R$ thermalization occurs at $z<1$, as shown in Fig.~\ref{fig:washout_2}. We conclude that when all the three $\nu_R$ are out of equilibrium at $T\sim m_\phi$, forbidden neutrinogenesis can only generate a maximum of the  $\nu_R$ asymmetry at $\mathcal{O}(10^{-11})$.

\subsubsection{One nonthermal neutrino}\label{sec:Y1}

When the two heavier $\nu_R$ have   larger Yukawa couplings such that they have reached thermal equilibrium during neutrinogenesis, a net $\nu_R$ asymmetry will only be accumulated in the lightest $\nu_R$ sector.  In this case, the nonthermal condition is provided by the scalar  and the lightest right-handed neutrino.   The calculation of the CP-violating source still comes from Fig.~\ref{fig:2loop_CP}, with the inner right-handed neutrinos (scalar) being in thermal equilibrium (out of equilibrium).  Following the general CP-violating source   derived in Ref.~\cite{Kanemura:2024dqv}, we arrive at  
\begin{align}\label{eq:S_CP_fin-2}
	\mathcal{S}_{\rm CP}=\frac{\text{Im}(y_4)m_\phi^4}{256\pi^4 (\tilde{m}_j^2-\tilde{m}_i^2)}&\int^\infty_0 \frac{dp_\ell}{p_\ell}\int^\infty_{\frac{m_\phi^2}{4p_\ell}+p_\ell} dE_\phi  \int^\infty_{\frac{m_\phi^2}{4p_\ell}+p_\ell} dE'_\phi I(p_\ell, E_\phi, E'_\phi)\,,
\end{align}
with $E'_\phi$ the  energy from the inner-loop scalar. Different from Eq.~\eqref{eq:statistics},  the   statistics function now reads
\begin{align}\label{eq:statistics-2}
	I(p_\ell, E_\phi, E'_\phi)=f^{\rm eq}_\phi(E_\phi) \delta f_\phi(E'_\phi)  \left[f_{\nu_R}^{\rm eq}(E'_\phi-p_\ell)+f^{\rm eq}_\ell(p_\ell)-1\right],
\end{align}
where $f_{\nu_R}^{\rm eq}$ is the thermal distribution function  for the two heavier $\nu_R$, and $\delta f_\phi=f_\phi-f_\phi^{\rm eq}$. Note that the dependence on   flavor $\beta$ disappears in the statistics function,  but we still have  $\sum_{\beta}\text{Im}(y_4)\neq 0$ since $\alpha$ is no longer a dummy index and is  fixed to be the lightest neutrino flavor, where 
\begin{align}\label{eq:Ynu1}
	Y_{\nu}= Y_{\nu_{R1}} =\int_{T_{\rm sph}}^\infty \frac{\mathcal{S}_{\rm CP}}{s_{\rm SM} \mathcal{H} T} dT\,.
\end{align}

\begin{table}
	\centering
	\renewcommand{\arraystretch}{1.4}
	\begin{tabular}{l|c|c}
		\hline\hline
		Patterns &  ($m_1,m_2,m_3$) (meV) & $10^{12}\mathcal{S}_{\rm CP}/\sin\delta_{\rm CP}$~($\rm eV^4$)\\ 
		\hline 
		NO with $m_1/v_2=10^{-6}$           &  	 ($10^{-3}$\,,\, 8.65\,,\, 50.13 )    &  $1.15 |\langle \delta f_\phi\rangle|$  \\
		\hline
		IO with $m_3/v_2=10^{-6}$     &  	  (49.08\,,\, 49.84\,,\, $10^{-3}$)  &  $0.04 |\langle \delta f_\phi\rangle|$  \\
		\hline	\hline
	\end{tabular}
	\caption{Behavior of $\mathcal{S}_{\rm CP}$ in the case of one nonthermal $\nu_R$,  where patterns NO+$m_{1,\rm max}$ and IO+$m_{3,\rm max}$ shown in Tab.~\ref{tab:neu-info} cannot realize forbidden neutrinogenesis.     $\mathcal{S}_{\rm CP}$  is estimated via Eq.~\eqref{eq:S_CP_est-2} with $|\langle \delta f_\phi\rangle|=-\langle \delta f_\phi\rangle$, and  $v_2=1$~eV is chosen.   }
	\label{tab:neu-info-2}
\end{table}

Analogously to Eq.~\eqref{eq:S_CP_est}, let us analyze the dependence on the PMNS phase and the neutrino mass spectrum. We now define 
\begin{align}\label{eq:S_CP_est-2}
	\mathcal{S}_{\rm CP}\equiv \frac{\text{Im}(y_4)}{8\pi^4 y_\mu^2}\langle \delta f_\phi\rangle\,,
\end{align}
where $\langle \delta f_\phi \rangle$\footnote{Note that $\langle \delta f_\phi \rangle<0$ is expected since $f_{\nu_R}^{\rm eq}+f^{\rm eq}_\ell-1\leqslant0$ and $\delta f_\phi>0$ is caused by the scalar mass effect, as indicated by Eq.~\eqref{eq:ddeltaf/dz}.} is a dimensionful  average of $\delta f_\phi$ under the 3-dimensional   integration of Eq.~\eqref{eq:S_CP_fin-2}.   We 
show in Tab.~\ref{tab:neu-info-2} the CP-violating rate in pattern	NO (IO) with $m_1/v_2=10^{-6}(m_3/v_2=10^{-6})$. Note that  the patterns with  $m_{1,\rm max}$ and $m_{3,\rm max}$ cannot realize neutrinogenesis since the lightest $\nu_R$ must be out of equilibrium. In contrast to the case of three nonthermal $\nu_R$, a positive $Y_B$ requires $0<\delta_{\rm CP}<\pi$ in both NO and IO patterns. This is expected since the dependence of  $\mathcal{S}_{\rm CP}$ on $\nu_R$ flavors now comes from $\text{Im}(y_4)$, which has the same dependence on $\delta_{\rm CP}$ in both NO and IO patterns.   Given that $0<\delta_{\rm CP}<\pi$ is disfavored in the IO pattern,\footnote{Both 
	the T2K  and NO$\nu$A experiments are consistent with $\pi<\delta_{\rm CP}<2\pi$  in the IO pattern within $3\sigma$ level. } we conclude that only the NO pattern can realize forbidden neutrinogenesis in the case of  one  nonthermal $\nu_R$.

 Comparing to the  case of three nonthermal $\nu_R$ that features small Yukawa couplings,  large Yukawa couplings  from the two heavier $\nu_R$   in the case of one nonthermal $\nu_R$ will enhance the CP-violating source. Such an enhancement plays a significant role in compensating   for   the    suppression  from a quasi-thermal scalar, where $\delta f_\phi$ is small. In general, there is larger parameter space to realize forbidden neutrinogenesis  in the case of one nonthermal $\nu_R$, depending on the scale of $|y_2|, |y_3|$. Here,  we will  consider a simple  situation where  both $|y_2|$ and $|y_3|$ are large enough to dominate the scalar evolution but small enough such that the resonant enhancement from   soft-lepton resummation is valid.\footnote{
We will demonstrate this point in Appendix~\ref{append:CTP-2}.}  In this case, the evolution of the scalar under the KB (or equivalently Boltzmann) equation with collision rates from decay and inverse decay is given by~\cite{Kanemura:2024dqv}
	\begin{align}\label{eq:fphi}
	\frac{\partial f_\phi}{\partial t}-\mathcal{H} p_\phi \frac{\partial f_\phi}{\partial p_\phi}	=-\delta f_\phi	\frac{|y_{\alpha}|^2 m_\phi^2T}{8\pi E_\phi p_\phi}\ln\left(\frac{\cosh \left(\frac{E_\phi+p_\phi}{4T}\right)}{\cosh \left(\frac{E_\phi-p_\phi}{4T}\right)}\right)\,,
\end{align}
where $p_\phi\equiv |\vec p_\phi|$, and $\alpha$ contains the two heavier $\nu_R$ flavors.

\begin{figure}[t]
	\centering
		\includegraphics[scale=0.29]{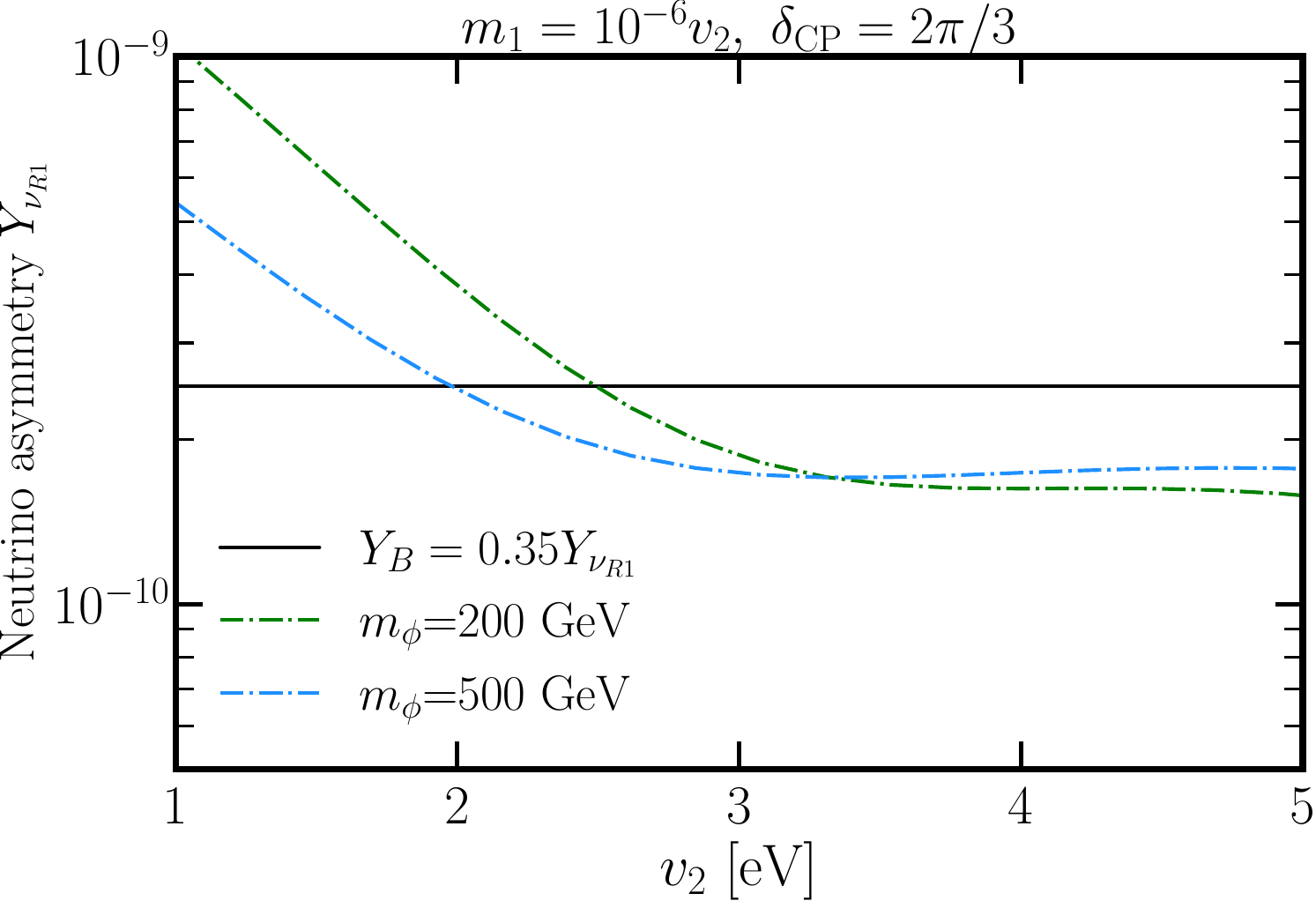}\quad
	\includegraphics[scale=0.29]{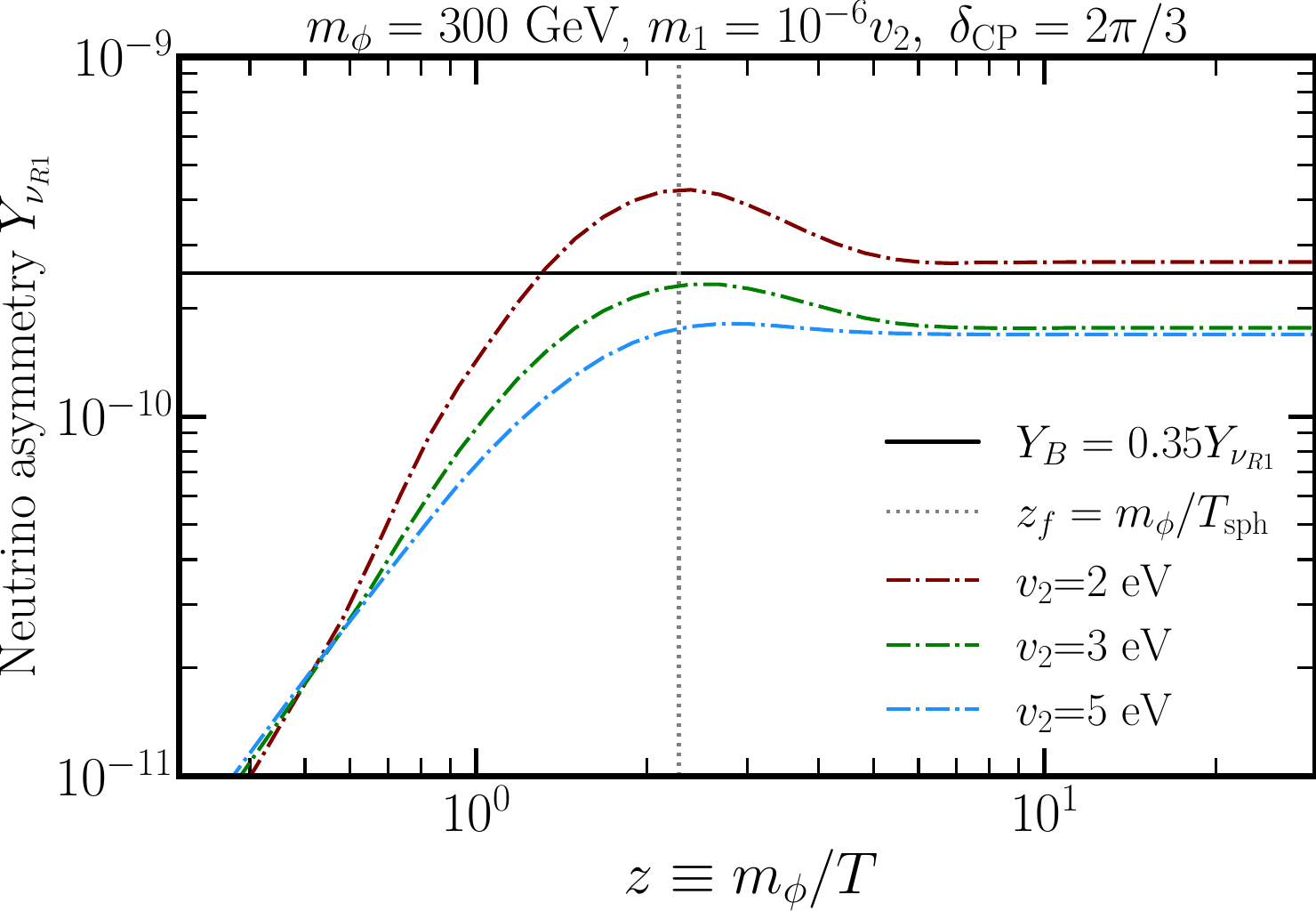}
	\caption{\label{fig:Y1} Left: the  magnitude of the  $\nu_{R1}$ asymmetry $Y_{\nu_{R1}}$ in terms of  $v_2$. Right: the evolution of $Y_{\nu_{R1}}$ from $m_\phi=300$~GeV, where the vertical dotted line denotes the epoch of sphaleron decoupling. Only the NO pattern works to generate a positive $Y_B$, where the Dirac CP-violating phase is fixed by $\delta_{\rm CP}=2\pi/3$.}
\end{figure}

Similarly to Fig.~\ref{fig:Y3}, we first present the scale of $v_2$ that can induce a large $\nu_{R1}$ asymmetry, which is shown in the left panel of Fig.~\ref{fig:Y1}. We fix $m_1/v_2=10^{-6}$   to avoid the washout effect, and the integration of temperature is cut in the IR regime via $z=m_\phi/T_{\rm sph}$. We see that $Y_{\nu_{R1}}$ at the order of $Y_B$ typically requires $v_2=\mathcal{O}(1)$~eV, which implies $|y_2|\sim |y_3|\sim 0.01$ for the two heavier $\nu_R$.   We see from the left panel that a lighter scalar results in a larger $Y_{\nu_{R1}}$ when $v_2$ is lower,  but a heavier scalar leads to a   larger $Y_{\nu_{R1}}$ when $v_2$ becomes   higher.
This can be  explained by the behavior of $\delta f_\phi$ from Eq.~\eqref{eq:fphi}, which has the structure
\begin{align}\label{eq:ddeltaf/dz}
	\frac{d\delta f_\phi}{dz}\sim -|y_\alpha|^2 \frac{M_P}{m_\phi} \delta f_\phi +\left|\frac{df_\phi^{\rm eq}}{dz}\right|.
\end{align}
We can see that the mass effect in $f_\phi^{\rm eq}$ drives the scalar into the out-of-equilibrium regime, which is the source for $\delta f_\phi$.  If there is no vacuum mass for the scalar before gauge symmetry breaking, i.e., $\mu_2\ll v_1\approx 246$~GeV, we would obtain  $df_\phi^{\rm eq}/dz=0$ and $\delta f_\phi=0$ will be maintained   from an initial equilibrium state. 
The first term on the right-hand side can be regarded as the exponential dilution  for $\delta f_\phi$. For larger $v_2$, the Yukawa couplings become smaller  and hence  the dilution will  become exponentially suppressed, increasing $\delta f_\phi$ thereby.  Note that this effect would be stronger for   larger scalar masses due to the ratio $M_{P}/m_\phi$.  The   resulting  enhancement of $\delta f_\phi$ can compete with    the power suppression from $\text{Im}(y_4)$, potentially allowing  a larger $Y_{\nu_{R1}}$ for higher $v_2$.   This explains the tendency   in the left panel of Fig.~\ref{fig:Y1} for $m_\phi=500$~GeV. It implies that a larger $v_2$ (smaller neutrino Yukawa couplings)  does not always lead to  $Y_{\nu_{R1}}$ suppression.

To see the evolution of $Y_{\nu_{R1}}$ in terms of the time variable $z\equiv m_\phi/T$, we show in the right panel of Fig.~\ref{fig:Y1} with a benchmark scalar mass: $m_\phi$=300~GeV. The ratio $m_1/v_2=10^{-6}$ is fixed to avoid the strong washout effect, and this can be justified from Fig.~\ref{fig:f1}, where $\nu_R$ thermalization occurs    after sphaleron decoupling. 
We set an initial condition at $z_i=10^{-3}$ with a thermal scalar and two thermal $\nu_R$.  In general, neutrinogenesis culminates at $z\sim 3$--5 due to Boltzmann suppression in the statistics function $I(p_\ell, E_\phi, E'_\phi)$.  However, a light scalar   can lead to earlier completion of neutrinogenesis due to sphaleron decoupling. In such a situation, neutrinogenesis ends before a final stable value of  $Y_{\nu_{R1}}$ is reached. This can be seen    through  the vertical dotted line ($z_f=m_\phi/T_{\rm sph}$) in the right panel for $m_\phi=300$~GeV.

We conclude that an electroweak scalar  is able to realize forbidden neutrinogenesis, where the neutrinophilic vacuum expectation value resides in  the eV scale. 
Forbidden neutrinogenesis  works to explain the BAU only in  the NO pattern with $\sin\delta_{\rm CP}>0$, where the lightest  $\nu_R$ is out of equilibrium during neutrinogenesis. This scenario is  highly falsifiable in upcoming neutrino oscillation experiments. In particular, 
 if  $\pi<\delta_{\rm CP}<2\pi$ turns out to be the truth for an NO neutrino mass spectrum: $m_1\ll m_2<m_3$,  it would be  able to  exclude the minimal neutrinophilic scalar scenario with  $v_2=\mathcal{O}(1)$~eV, as the framework   can readily induce a large negative baryon asymmetry in the early universe via forbidden neutrinogenesis.

 \section{Discussions}\label{sec:dis}

\subsection{Comparison with the Boltzmann equation}\label{sec:KBvsBol}
Forbidden leptogenesis  via lepton-number  conserving  interactions was  previously  considered  in Refs.~\cite{Li:2020ner,Li:2021tlv}, where  the Boltzmann equation was used  and the kinetic phase was derived  from the retarded/advanced cutting rules in the one-loop self-energy diagrams.  Moreover, the nonthermal condition that realized    forbidden leptogenesis was only provided by one $\nu_R$  flavor.   This is qualitatively inconsistent  with the results obtained from the SK-CTP formalism followed by the KB equation~\cite{Kanemura:2024dqv}, which, as also shown in previous sections,  points out that    additional nonthermal conditions must be provided either by the scalar or by the other $\nu_R$ flavors from the inner loop of Fig.~\ref{fig:2loop_CP}. 

 The reason behind such   inconsistency is that  the evolution of CP asymmetries generated by  the pure plasma effect was determined  by the  Boltzmann equation in  Refs.~\cite{Li:2020ner,Li:2021tlv}, where the   real-intermediate-state subtraction~\cite{Kolb:1979qa}  was not taken properly. It is known that  if the on-shell scattering effect in the Boltzmann equation was not included consistently,  CP asymmetries can still be generated even in thermal equilibrium. This is   inconsistent with unitarity and the CPT theorem~\cite{Giudice:2003jh}.  In Refs.~\cite{Li:2020ner,Li:2021tlv}, only the decay and inverse decay processes were included in the Boltzmann  equation, which neglected the on-shell scattering process that has a canceling effect on the CP asymmetry induced by decay/inverse decay. 

Real-intermediate-state subtraction   should also be taken into account  in forbidden leptogenesis, which, however,  can be circumvented   in the SK-CTP formalism in a straightforward way~\cite{Beneke:2010wd,Kanemura:2024dqv}.  To see this, let us recall the five contributions to the CP-violating source $\mathcal{S}_{\rm CP}$, which are parameterized by  functions $\mathcal{I}_i$ and $\mathcal{J}_i$ given in Appendix~\ref{append:SCP_EWscalar}. The contributions of decay and inverse decay  arise from $\mathcal{I}_{1,3,4,5}$ and $\mathcal{J}_{1,3,4,5}$, while  the on-shell scattering contributions would arise   from $\mathcal{I}_{2}$ (Eq.~\eqref{eq:I2}) and $\mathcal{J}_{2}$ (Eq.~\eqref{eq:J2}) when one of the resummed  retarded lepton  propagators $\slashed{{S}}^R_{\ell_i}$ or $\slashed{{S}}^R_{\ell_j}$ goes on-shell. Such on-shell scattering contributions  ensure the appearance of the \textit{thermal-criterion} function $\mathcal{TC}$ given in Eq.~\eqref{eq:TC}, which  guarantees no CP asymmetry in thermal equilibrium (Eq.~\eqref{eq:KMS}) and hence     consistency with unitarity and the CPT theorem. 

It is not  clear yet whether one can start from the Boltzmann equation with one-loop self-energy amplitudes to evaluate forbidden leptogenesis. The reason behind is that there is no simple principle to write down the  propagators in the  one-loop self-energy diagrams built in  the Boltzmann collision rates. In particular, inserting  the thermal mass into fermion  Feynman  propagators  is not justified theoretically.  Cutting rules, or circling rules at finite-temperature field theory~\cite{Kobes:1985kc,Kobes:1986za} provide a possibility for studying purely plasma-induced effects within the Boltzmann equation, but are as technically nontrivial as the calculation of the KB collision rates in the   SK-CTP formalism. Finding a simple correspondence between the two approaches deserves further  considerations.

 \subsection{Phenomenology}\label{sec:pheno}
 The minimal neutrinophilic scalar scenario introduces several new-physics effects that could  be observed in cosmology, low-energy flavor physics and colliders.  In this section, we will briefly discuss some of the  potential signals that follow   the realization of forbidden neutrinogenesis. 
 
  In cosmology,  
 since three $\nu_R$ are the  right-handed Dirac counterparts of the SM $\nu_L$, they contribute as extra radiation to the expansion of the universe, affecting processes  during the big-bang nucleosynthesis (BBN) and cosmic microwave background (CMB) epochs.   After neutrino decoupling, the effect of the accelerated cosmic  expansion due to  relativistic $\nu_R$    is     parameterized by the effective neutrino number:
\begin{align}\label{eq:Neff_def}
	\Delta N_{\rm eff}=	\sum_\alpha \Delta N_{\alpha, \rm eff}\equiv \sum_\alpha  \frac{\rho_{\alpha}}{\rho_\nu^{\rm SM}}\,,
\end{align} 
where $\rho_\alpha$ is the energy density of $\nu_{R\alpha}$, and $\rho_{\nu}^{\rm SM}$ denotes  the energy density of one-generation $\nu_L$  in the SM:
\begin{align}
	\rho_{\nu}^{\rm SM}=\frac{7}{4}\frac{\pi^2}{30}T_\nu^4\,,
\end{align}
with  $T_\nu\approx (4/11)^{1/3}T$ after neutrino decoupling. 

Thermalized  right-handed neutrinos, as predicted in the case of one nonthermal $\nu_R$,  will give  significant contributions  to $\Delta N_{\rm eff}$.  Using Eq.~\eqref{eq:Neff_def} and entropy conservation, one can easily derive 
\begin{align}\label{eq:Delat_N_alpha_eq}
	\Delta N_{\alpha, \rm eff}
	\approx 0.027\left(\frac{106.75}{g_s(T_{\alpha})}\right)^{4/3}g_{\alpha}\,,
\end{align}
for thermalized $\nu_R$ of flavor $\alpha$. Here, $g_s(T_{\alpha})$ is the SM effective degrees of freedom  at the $\nu_{R\alpha}$ decoupling temperature  $T_{\alpha}$,  and we have taken a reference point $g_s=106.75$ from relativistic   SM species at $T=\mathcal{O}(100)$~GeV. 
$g_{\alpha}=2\times 7/8$ is the effective spin degrees of freedom for  $\nu_{R\alpha}$.

\begin{figure}[t]
	\centering
	\includegraphics[scale=0.55]{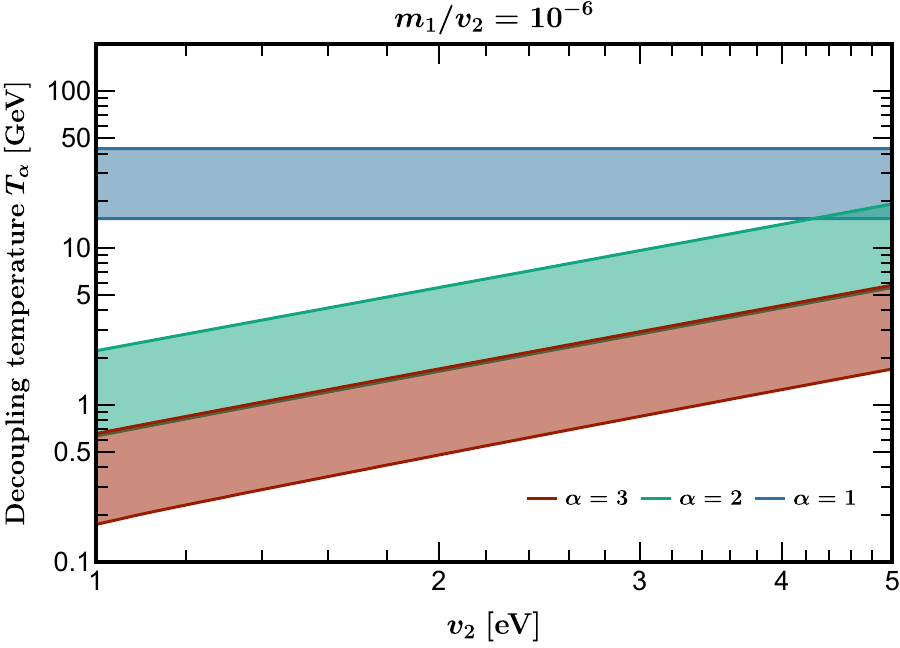} 
	\caption{\label{fig:Tdec} The decoupling temperatures of the three  $\nu_R$ flavors in the NO pattern, where $\alpha=3 (1)$ corresponds to the heaviest (lightest) $\nu_R$. The upper (lower)  boundary for each $\alpha$ corresponds to $m_\phi=500\,(200)$~GeV. We set $m_1/v_2=10^{-6}$ such that $T_1$ is independent of $v_2$.  }
\end{figure}

As elaborated in section~\ref{sec:th-cal}, only the case of one nonthermal $\nu_R$ can realize forbidden neutrinogenesis to explain the BAU, where the NO pattern with $0<\delta_{\rm CP}<\pi$ is favored. In this case, the  two heavier $\nu_R$ flavors have already reached thermal equilibrium at the neutrinogenesis epoch, while the lightest $\nu_R$ flavor will also establish thermalization after sphaleron decoupling.  Nevertheless,   the late-time evolution is  not the same for the three $\nu_R$. 
 Due to the large Yukawa couplings ($|y_2|\sim |y_3|\sim 0.01$),  right-handed neutrino scattering  with charged leptons can   maintain thermal equilibrium at $T\ll m_\phi$, and  thermal equilibrium with  $\nu_L$ could  persist even after $T=\mathcal{O}(1)$~GeV.  For the lightest $\nu_R$,   the much smaller Yukawa coupling ($|y_1|\sim 10^{-6}$) renders a suppressed four-fermion scattering rate such that  left-right neutrino scattering will not maintain the   equilibrium condition.  Instead, nonrelativistic scalar decay and inverse decay $\phi \leftrightharpoons \ell+\bar \nu_R$ will determine the decoupling temperature of the lightest $\nu_R$.
 
  To  determine the decoupling temperature $T_{\alpha}$ for $\nu_{R2}$ and $\nu_{R3}$,  we  assume  the neutral scalar boson $H$    is the lightest one in   the neutrinophilic scalar     doublet,  and focus  on the $t$-channel $\nu_R+\bar \nu_R\leftrightharpoons  \nu_L +\bar \nu_L$ mediated by $H$. The annihilation cross section     reads
\begin{align}
\sigma_{\nu_R\bar \nu_R \to \nu_L \bar \nu_L}=\frac{m_i^2 m_j^2 s}{24\pi (m_Hv_2)^4}\,,
\end{align}
where $\sqrt{s}$ is the center-of-mass energy and the physical mass $m_H$  is given in Eq.~\eqref{eq:Higgsmass}.  The appearance of the two heavier neutrino masses $m_i, m_j$   results from the application of Eq.~\eqref{eq:yalpha_sq}. We substitute the annihilation cross section    into the Boltzmann equation   and obtain the thermally averaged annihilation rate~\cite{Gondolo:1990dk}:
\begin{align}\label{eq:sigmavn}
	\langle \sigma v n\rangle \equiv \frac{T}{32\pi^4 n_{\nu_R}^{\rm eq}}\int_0^{\infty} ds \sigma_{\nu_R\bar \nu_R \to \nu_L \bar \nu_L} s^{3/2} K_1(\sqrt{s}/T)\,,
\end{align}
where  $K_1$ is the modified Bessel function of the second kind, and  $n_{\nu_R}^{\rm eq}$ is the  number density of thermalized $\nu_R$. 
For the decoupling temperature of the lightest $\nu_R$, we again concentrate on the neutral scalar boson  $H$ with the decay rate 
\begin{align}
	\Gamma_{H\to \nu_L\bar \nu_R}=\frac{m_1^2}{16\pi v_2^2} m_H\,,
\end{align}
and with the nonrelativistic number density 
\begin{align}
	n^{\rm eq}_H=\left(\frac{m_H T}{2\pi}\right)^{3/2} e^{-m_H/T}\,.
\end{align}
We determine the decoupling temperatures $T_{i}$  via\footnote{While we   do not intend to calculate the precise $T_{\alpha}$ here, we should mention that this treatment, together with Eq.~\eqref{eq:sigmavn} that is valid by taking the approximation of the Boltzmann distribution, leads to some uncertainty in the $T_{\alpha}$ calculation. Nevertheless, Eq.~\eqref{eq:Delat_N_alpha_eq} suggests that  a precise $T_{\alpha}$ only becomes significant around the QCD phase transition $T_{\rm QCD}\sim 200$~MeV.  }
\begin{align}
\mathcal{H}&=\langle \sigma v n\rangle\,, \quad \text{for}~~T_2\,, T_3\,, 
\\[0.2cm]
n_{\nu_R}^{\rm eq}\mathcal{H}&=n_H^{\rm eq} \Gamma_{H \to \bar \nu_L \nu_R}\,, \quad \text{for}~~T_1\,.
\end{align}
Motivated by the realization of forbidden neutrinogenesis in the case of one nonthermal $\nu_R$, we choose the parameter set: $v_2=[1,5]$~eV, $m_1/v_2=10^{-6}$ and $m_\phi=200,500$~GeV, and show the decoupling temperatures in  Fig.~\ref{fig:Tdec}.  At the lower $m_H\cdot v_2$ end,  i.e., $m_H=200$~GeV and $v_2=1$~eV, the decoupling temperatures $T_i$ are found to be around $170, 630, 1540$~MeV, respectively. This  amounts to $\Delta N_{\rm eff}\approx 0.377$, which is   larger than the current Planck bound~\cite{Planck:2018vyg}: $\Delta N_{\rm eff}<0.285$.  At the higher $m_H\cdot v_2$ end,  with  $m_H=500$~GeV and $v_2=5$~eV,  on the other hand, the decoupling temperatures are found to be $5.8, 19, 43$~GeV, respectively,  giving  rise to $\Delta N_{\rm eff}\approx 0.186$. This lower value will be covered by future sensitivity from \textit{e.g.,} the Simons Observatory experiment~\cite{SimonsObservatory:2018koc}. Therefore, cosmic measurements of $N_{\rm eff}$ will  provide a robust  test for  forbidden neutrinogenesis in the neutrinophilic scalar scenario.

Next, let us consider the signatures from low-energy flavor physics.  It has been noticed that one of the most sensitive probes for the neutrinophilic scalar scenario comes from lepton-flavor violating transitions~\cite{Bertuzzo:2015ada}, particular in    $\mu\to e\gamma$,  $\mu\to 3e$, and $\mu\to e$ in nuclei. Current bounds   exclude the  charged-scalar mass below 250~GeV for $v_2=1$~eV.\footnote{We  mention that the charged-scalar mass given in Eq.~\eqref{eq:Higgsmass} can be larger than   the mass ($\mu_2$)  used in neutrinogenesis before gauge symmetry breaking.} Given that  future sensitivities are forecast to increase by one to several orders of magnitude, such as the COMET experiment from J-PARC~\cite{COMET:2018auw}, MEG-II~\cite{MEGII:2018kmf} and Mu3e~\cite{Blondel:2013ia}, we expect that  an  electroweak neutrinophilic scalar with $v_2= \mathcal{O}(1)$~eV can be fully tested.  Moreover,  the new-physics effects on these lepton-flavor violating transitions are correlated. Taking $\mu\to e\gamma$ and   $\mu\to 3e$ for example, we have the branching ratios~\cite{Bertuzzo:2015ada,Li:2022yna}:
\begin{align}
\frac{\mathcal{B}(\mu\to 3e)}{\mathcal{B}(\mu\to e\gamma)}\approx \frac{\alpha_{\rm EM}}{36\pi}\left[24\ln\left(\frac{m_\mu}{m_e}\right)-43\right],
\end{align}
where the approximation is valid  in the regime  $v_2\gtrsim1$~eV and takes into account the fact that charged leptons and Dirac neutrinos  are much lighter than the charged scalar  bosons. It is worth mentioning that in scenarios with heavy right-handed (Majorana) neutrinos coupling to  scalar doublets, there could also be simple correlations between $\mu\to e\gamma$ and   $\mu\to 3e$, such as the  supersymmetric low-scale seesaw model~\cite{Arganda:2005ji,Ilakovac:2012sh}, the Tao-Ma scotogenic model~\cite{Tao:1996vb,Ma:2006km} and the AKS model~\cite{Aoki:2008av,Aoki:2009vf}. Unlike the minimal neutrinophilic scalar scenario with Dirac neutrinos,   the correlations induced by  these scenarios  usually depend  on additional Yukawa couplings and on the mass spectrum between the heavy Majorana neutrinos and the charged scalar bosons~\cite{Aoki:2011zg,Toma:2013zsa,Enomoto:2024jyc,Kanemura:2024aja}.

  \begin{figure}[t]
  	\centering
  	\includegraphics[scale=0.55]{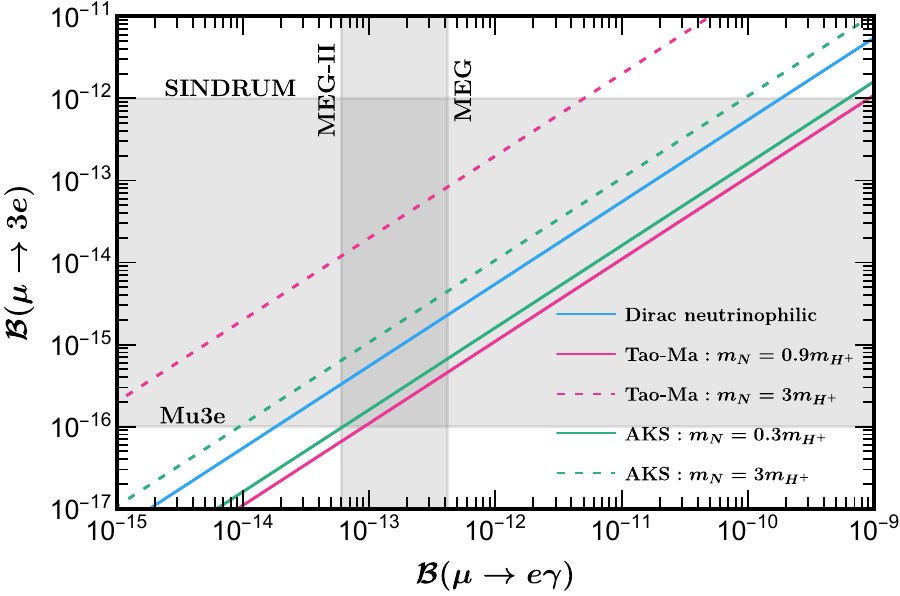} 
  	\caption{\label{fig:LFV} The correlation between the branching ratios of $\mathcal{B}(\mu\to 3e)$ and $\mathcal{B}(\mu\to e\gamma)$ in the Dirac neutrinophilic scalar scenario and some typical models with Majorana neutrinos.  Current and future bounds are shown in shaded regions. The predictions from  the Tao-Ma~\cite{Tao:1996vb,Ma:2006km} and the AKS~\cite{Aoki:2008av,Aoki:2009vf} models are shown with different mass ratios between the heavy Majorana neutrinos and charged scalar bosons, where order-one Yukawa couplings are assumed.}
  \end{figure}

We show  in Fig.~\ref{fig:LFV} the correlation between   $\mathcal{B}(\mu\to 3e)$ and $\mathcal{B}(\mu\to e\gamma)$, with   the current (future) bounds of $\mu\to e\gamma$ from MEG~\cite{MEG:2016leq} (MEG-II~\cite{MEGII:2018kmf}) and of $\mu\to3e$ from SINDRUM~\cite{SINDRUM:1987nra} (Mu3e~\cite{Blondel:2013ia}).  As a comparison with  the Dirac neutrinophilic scalar scenario, we also show the predictions from the Tao-Ma model and   the AKS model with different mass ratios ($m_N/m_{H^+}$) between the heavy Majorana neutrinos and the charged scalar bosons. Notice that these predictions usually depend on additional Yukawa couplings, so we take order-one Yukawa couplings  for illustration. It is seen that  the predicted correlation line  from the Dirac neutrinophilic scalar scenario lies between the two exemplary mass ratios in the  Tao-Ma  model and  the AKS model. It implies that there exists certain degeneracy among these models in some parameter space, but such degeneracy can usually  be removed via complementary probes. 

In the Dirac neutrinophilic scalar scenario, the branching ratios of   $\mu\to e\gamma$ and   $\mu\to 3e$ depend on the product of $m_{H^\pm}$ and   $v_2$~\cite{Bertuzzo:2015ada,Li:2022yna}. With increased sensitivities  from the future MEG-II experiment  on $\mu\to e\gamma$  and  from the future Mu3e experiment on $\mu\to 3e$ , we find that the correlated region that can be  probed simultaneously by   MEG-II and  Mu3e    is induced  by $m_{H^\pm}\cdot v_2=[332,540]~\text{GeV}\cdot \text{eV}$. For example, a charged-scalar mass at 300~GeV with $v_2=1.8$~eV can be  probed via the correlated   $\mu\to e\gamma$ and   $\mu\to 3e$. This region covers the parameter space that can realize  forbidden neutrinogenesis presented in section~\ref{sec:Y1}.  The simultaneous observation of $\mu\to 3e$ and $\mu\to e\gamma$ in the future would provide an interesting  indication of such a minimal neutrinophilic scalar scenario that can explain the BAU and neutrino masses. 

Finally, we would like to comment on    collider detection. For the minimal neutrinophilic scalar scenario that can realize the forbidden neutrinogenesis, there are several collider detection channels for the scalar bosons.  For electroweak charged-scalar bosons,  gauge interactions allow sizable cross sections from Drell-Yan production: $p p\to Z^*/\gamma^*\to H^+ H^-$ at the  LHC~\cite{Davidson:2010sf,Akeroyd:2016ymd},  which is   followed  by prompt decay   $H^\pm \to \ell^{\pm}+\nu_R$ due to large  neutrino Yukawa couplings. Mono-lepton signals can arise from
production of  neutral  and charged scalars, e.g.,  $p  p\to W^{\pm *} \to H^\pm A$~\cite{Kanemura:2001hz,Cao:2003tr}  followed by $H^\pm \to \ell^{\pm}+\nu_R$ and $A \to \nu_L+\bar \nu_R$.   Purely neutral-scalar production can be constructed via $p  p \to Z^*\to H A$, which, however, will be mostly followed by prompt decay to neutrinos, again due to large  neutrino Yukawa couplings. Future lepton colliders, such as the Compact Linear Collider~\cite{Linssen:2012hp,CLIC:2018fvx}, the International Linear Collider~\cite{ILC:2007bjz,ILC:2013jhg} and the muon collider~\cite{Delahaye:2019omf,Accettura:2023ked} will also be useful to produce these new scalars. In particular,  the mono-photon channel from initial-state radiation $e^+ e^- \to Z^*\to H A (\to \text{missing~energy})+\gamma$ can provide an interesting avenue for purely neutral-scalar production, where the missing four-momentum can be calculated by using the photon energy and the photon polar angle~\cite{Birkedal:2004xn,Bartels:2012ex}.

 It has been a haunting concern to test leptogenesis at colliders, where  generally no   smoking-gun signal can be uniquely induced  from  leptogenesis. This situation would become worse if leptogenesis is realized  by new physics with flavored scalars, since  the indication of the BAU origin requires confirmation of the scalar family. Here, we would like to emphasize  an \textit{inverse perspective} of  forbidden leptogenesis in motivating collider detection: if some new particle is detected at colliders, can this particle be ready to resolve the BAU problem? As elaborated in this paper,  the BAU origin can be explained   if  there is an electroweak  neutrinophilic scalar minimally introduced beyond  the SM. While the collider signals mentioned above may not be the unique smoking-gun, 
 they   provide a  simple indication that the BAU can already be  attributed to such minimal  new physics.  This is  different from flavored-scalar leptogenesis~\cite{Covi:1996fm,Ma:1998dx,Dick:1999je,Murayama:2002je}, where  more delicate collider confirmation is needed.

 \section{Conclusion}\label{sec:con}
 Plasma effects via soft-lepton resummation exhibit a     resonant  enhancement in generation of finite-temperature CP asymmetries, which can reach  $\mathcal{O}(10^{8})$ from quasi-degenerate lepton thermal masses.  The enhancement is  predicted within the SM and does not require vacuum mass degeneracy.   Applying this  kind of forbidden leptogenesis  to the neutrinophilic scalar scenario in the SK-CTP formalism, we have elaborated that  the minimal framework for the  Dirac neutrino mass origin   can further explain the BAU problem, where the scalar and one right-handed neutrino provide the nonthermal condition during neutrinogenesis. 
 
The CP-violating source for forbidden neutrinogenesis in the minimal  scenario comes directly from neutrino mixing. The Dirac CP-violating phase   ($\delta_{\rm CP}$)  in the PMNS matrix then determines the sign of the baryon asymmetry. We find that the BAU explanation only favors the  normal-ordering neutrino mass spectrum, where $\delta_{\rm CP}$ should be  in $(0,\pi)$ and the lightest neutrino is predicted to have a much smaller mass $m_1\ll m_2\approx  m_3/6$. The connection between high-temperature neutrinogenesis and low-energy CP violation  makes this scenario readily testable via upcoming measurements in neutrino oscillation experiments. In particular, either the  inverted or normal ordering with $m_1\ll m_2$  and $\pi<\delta_{\rm CP}<2\pi$ is able to  exclude the minimal scenario with an electroweak scalar and an   eV-scale vacuum expectation value, since the framework would  generate a large negative baryon asymmetry in the early universe. On the other hand,  forbidden neutrinogenesis cannot be realized in the minimal  scenario if $m_1\sim  m_2$ is confirmed. 
 
BBN and CMB measurements of $N_{\rm eff}$ in cosmology and the correlation between lepton-flavor violating transitions provide complementary probes of the minimal scenario, and will fully cover the parameter space for forbidden neutrinogenesis with forecast experimental  sensitivities.  Finally,  the electroweak scalar  accessible at colliders also provides another way to justify if the BAU problem can be already attributed to such  minimal new physics.  
  
The detailed  calculation presented in this work also helps to understand the behavior of  SM thermal leptons, with the aim of  exploiting the maximal role of particles  at finite temperatures. It  allows stronger connections among  complementary  probes, minimizing  the particle content in realizing  leptogenesis.

\section*{Acknowledgements}
  We would like to thank Kei Yagyu for helpful discussions on the collider phenomenology of neutrinophilic scalars. This project is supported by JSPS Grant-in-Aid for JSPS Research Fellows No. 24KF0060. SK is also supported in part by Grants-in-Aid for Scientific Research(KAKENHI) Nos. 23K17691 and 20H00160.

\appendix

\section{Propagators in the SK-CTP formalism}\label{append:CTP-1}
In the  CTP formalism,  if the system is close to thermal equilibrium, the free fermion propagators   in a spatially homogeneous plasma can be formulated by
\begin{align}\label{eq:S<}
i \slashed{S}^<(p)&=-2\pi \delta(p^2-m^2)(\slashed{p}+m)\left[\theta(p_0)f(p_0)-\theta(-p_0)(1-\bar f(-p_0))\right],
	\\[0.2cm]
i \slashed{S}^>(p)&=-2\pi \delta(p^2-m^2)(\slashed{p}+m)\left[-\theta(p_0)(1-f(p_0)+\theta(-p_0)\bar f(-p_0)\right],\label{eq:S>}
	\\[0.25cm]
i\slashed{S}^T(p)&=\frac{i(\slashed{p}+m)}{p^2-m^2+i\epsilon}-2\pi \delta(p^2-m^2)(\slashed{p}+m)\left[\theta(p_0)f(p_0)+\theta(-p_0)\bar f(-p_0)\right],\label{eq:ST}
	\\[0.25cm]
	i\slashed{S}^{\bar T}(p)&=-\frac{i(\slashed{p}+m)}{p^2-m^2-i\epsilon}-2\pi \delta(p^2-m^2)(\slashed{p}+m)\left[\theta(p_0)f(p_0)+\theta(-p_0)\bar f(-p_0)\right],\label{eq:STbar}
\end{align}
and for scalar bosons, 
\begin{align}\label{eq:freeProp-G}
 	i G^<(p)&=2\pi \delta(p^2-m^2)\left[\theta(p_0)f(p_0)+\theta(-p_0)(1+\bar f(-p_0))\right],
	\\[0.2cm]
 	i G^>(p)&=2\pi \delta(p^2-m^2)\left[\theta(p_0)(1+f(p_0))+\theta(-p_0)\bar f(-p_0)\right],
	\\[0.25cm]
 	iG^T(p)&=\frac{i}{p^2-m^2+i\epsilon}+2\pi \delta(p^2-m^2)\left[\theta(p_0)f(p_0)+\theta(-p_0)\bar f(-p_0)\right],
	\\[0.25cm]
	 	iG^{\bar T}(p)&=-\frac{i}{p^2-m^2-i\epsilon}+2\pi \delta(p^2-m^2)\left[\theta(p_0)f(p_0)+\theta(-p_0)\bar f(-p_0)\right],
\end{align}
where  $\lessgtr$   represent the Wightman functions and $T(\bar T)$ the (anti) time-ordered propagators. $\theta(x)$ denotes the Heaviside step function. The above formulation is based on the KB ansatz in the quasi-particle approximation, which is a justified approximation in  a close-to-equilibrium plasma. See \textit{e.g.} Refs.~\cite{Prokopec:2003pj,Berges:2004yj,Prokopec:2004ic}. Under this approximation, $f,\bar f$  are regarded as  the distribution functions of particles and antiparticles.
In thermal equilibrium,  we have
\begin{align}
	f^{\rm eq}(p_0)=\bar f ^{\rm eq}(p_0)=\frac{1}{e^{p_0/T}\pm 1}\,
\end{align}
for fermions ($+$) and bosons ($-$). Besides, the Kubo–Martin–Schwinger (KMS)  relations hold:
\begin{align}
\slashed{S}^>(p)=-e^{p_0/T}\slashed{S}^<(p)\,, \quad G^>(p)=e^{p_0/T}G^<(p)\,.
\end{align}

In general, for a system far away from thermal equilibrium, the above formulation could fail to describe the dynamics and evolution~\cite{Lindner:2007am,Anisimov:2008dz}. Fortunately, for most leptogenesis scenarios, the system is close to thermal equilibrium. This is also the case in forbidden neutrinogenesis, since  the dominant CP-asymmetry generation culminates at $0.1\lesssim T/m_\phi\lesssim 1$,  when both the scalar and right-handed neutrinos are close to    thermal equilibrium.

\section{Soft-lepton  resummation in Hard-Thermal-Loop approximation}\label{append:CTP-2}
At high temperatures, a (nearly) massless particle propagating  in a thermal plasma will receive significant correction from the plasma if the propagation momentum is lower than the plasma temperature. An intuitive way to see the importance is to consider a massless propagator $1/p^2$, which can suffer from IR divergence when $p\to 0$. However, one-loop self-energy diagrams at finite temperatures can give $g^2 T^2$ correction to the dispersion relation, where $g$ denotes a generic coupling in the theory.  It implies that soft-momentum propagation at $p\sim g T$ should be taken properly as the   IR enhancement    will appear at $p\sim g T$ instead of $p=0$. A technique  to properly resolve this soft-momentum propagation at finite temperatures is the so-called Hard-Thermal-Loop (HTL) resummation~\cite{Braaten:1989mz,Frenkel:1989br,Braaten:1991gm,Carrington:1997sq,Bellac2000}. In the HTL approximation, one-loop self-energy amplitudes are calculated by treating the external momentum at soft  scale $gT$ while taking the loop momentum at hard scale $T$. This has been  implemented  in   soft-lepton resummation when we calculate the CP-violating rate from Fig.~\ref{fig:2loop_CP}.

The resummed Wightman functions for SM leptons appeared  in Eqs.~\eqref{eq:tildeS<Lij}-\eqref{eq:tildeS>Lij} can be expressed in terms of the resummed retarded  $\slashed{{S}}^R$  and advanced $\slashed{{S}}^A$  propagators:
\begin{align}\label{eq:tilde-S<>}
	\slashed{S}^<(p)&=-f(p_0)\left[\slashed{S}^R(p)-\slashed{S}^A(p)\right],
	\\[0.2cm]
	\slashed{{S}}^>(p)&=\left[1-f(p_0)\right]\left[\slashed{{S}}^R(p)-\slashed{{S}}^A(p)\right],
\end{align}
with the following relations
\begin{align}\label{eq:S-relations}
	\slashed{{S}}^T- \slashed{{S}}^{\bar T}=\slashed{{S}}^R+\slashed{{S}}^A, \quad \slashed{{S}}^>- \slashed{{S}}^<=\slashed{{S}}^R-\slashed{{S}}^A\,.
\end{align}
The resummed retarded propagator    reads~\cite{Weldon:1982bn}
\begin{align}
	\slashed{{S}}^{R}(p)= \frac{(1+a_{})\slashed{p}+b_{}\slashed{u}}{\left[(1+a_{})p_0+b_{}\right]^{2}-\left[(1+a_{})|\vec{p}|\right]^{2}} 
\equiv \sum_{\pm}\frac{1}{\text{Re}\Delta_{\pm}+i \text{Im}\Delta_{\pm}}{P}_\pm\,,\label{eq:tildeS^R-def}
\end{align}
where $u_{\mu}$ is the four-velocity of the plasma normalized by $u_{\mu}u^{\mu}=1$  with $u_{\mu}=(1,0,0,0)$ in the rest frame. Note that the resummed advanced propagator can be obtained from retarded propagator via $a,b\to a^*, b^*$.  The     dispersion relation and  the thermal  width  of  leptons are determined by the real and imaginary parts, respectively:
\begin{align}
	\text{Re}\Delta_{\pm}(p)&\equiv(1+\text{Re} a_{})(p_{0}\pm |\vec{p}|)+\text{Re}b_{}\,,
	\\[0.2cm]
	\text{Im}\Delta_{\pm}(p)&\equiv \text{Im} a_{}(p_{0}\pm |\vec{p}|)+\text{Im}b_{}\,,
\end{align}
and  $P_\pm$ denotes the decomposition of helicity eigenstates~\cite{Braaten:1990wp}
\begin{align}
	P_\pm\equiv  \frac{\gamma^{0}\pm \vec{e}_{p}\cdot\vec{\gamma}}{2}\, ,
\end{align}
with $\vec{e}_p\equiv\vec{p}/|\vec{p}|$. 

For   thermal particles with vacuum masses much smaller than the plasma temperature, the real coefficients $\text{Re}a_{}, \text{Re}b_{}$ can be analytically derived   in the HTL approximation:
\begin{align}\label{eq:aRi}
	\text{Re}a_{i}&=\frac{\tilde{m}^2_{i}}{|\vec{p}|^2}\left[1+\frac{p_0}{2|\vec{p}|}\ln\left(\frac{p_0-|\vec{p}|}{p_0+|\vec{p}|}\right)\right],
	\\[0.3cm]
	\text{Re}b_{i}&=-\frac{\tilde{m}^2_{i}}{|\vec{p}|}\left[\frac{p_0}{|\vec{p}|}-\frac{1}{2}\left(1-\frac{p_0^2}{|\vec{p}|^2}\right)\ln\left(\frac{p_0-|\vec{p}|}{p_0+|\vec{p}|}\right)\right]\,, \label{eq:bRi}
\end{align}
 where $\tilde{m}$ denotes the lepton thermal mass in the SM:
\begin{align}\label{eq:thermalmass}
	\tilde{m}_{i}^2=\left(\frac{3}{32}g_2^2+\frac{1}{32}g_1^2+\frac{1}{16}y_{\ell_i}^2\right)T^2\,,
\end{align}
with $g_2, g_1$   the $SU(2)_L$ and $U(1)_Y$  gauge couplings, and $y_{\ell_i}$   the charged-lepton Yukawa couplings.  To   understand why $\tilde{m}_i$ works as the effective mass, we  express  the pole $\text{Re}\Delta_{-}=0$ as:
\begin{align}\label{eq:p0-p}
	p_0- |\vec p|=-\frac{\text{Re}b_i}{1+\text{Re}a_i}\approx \frac{\tilde{m}_i^2}{|\vec p|}=\frac{m_{\rm eff}^2}{p_0+|\vec p|}\sim \frac{m_{\rm eff}^2}{|\vec p|}\,,
\end{align}
where the first approximation is obtained by taking $1+\text{Re}a_i= \mathcal{O}(1)$ and using the leading-order relation $p_0/ |\vec p|\approx 1$ in the higher-order coefficient $\text{Re}b_i$. The effective mass is defined by $m_{\rm eff}^2\equiv p_0^2-|\vec p|^2$ from which we can see the correspondence $\tilde{m}\sim m_{\rm eff}$.  While the above simple derivation gives us a clear sense that $\tilde{m}_i$ works as an effective mass at finite temperatures, more precise computation from the  poles $\text{Re}\Delta_{\pm}=0$  shows that the solutions can be well  approximately  by two modes: $p_0^2-|\vec p|^2\approx 0$ and $p_0^2-|\vec p|^2\approx 2\tilde{m}_i^2$~\cite{Kiessig:2010pr,Drewes:2013iaa,Li:2023ewv}, which is valid for soft momentum $p_0\sim |\vec p|\sim gT$. The second mode, which is lepton-flavor dependent, leads to a nonzero CP-violating rate after summing the lepton flavors in Fig.~\ref{fig:2loop_CP}.

\begin{figure}[t]
	\centering
	\includegraphics[scale=0.7]{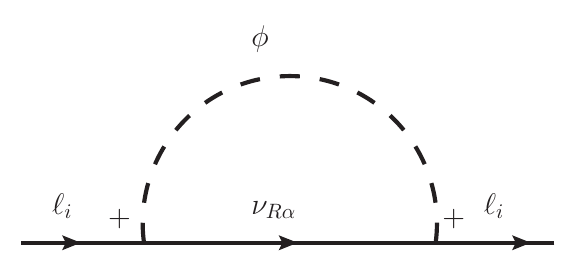} 
	\caption{\label{fig:tildem-NP} The new-physics contribution from Eq.~\eqref{eq:lag} to lepton thermal masses, which comes from   the real part of the time-ordered amplitude $\text{Re}\Sigma_\ell^{++}\equiv \text{Re}\Sigma_\ell^{T}$. }
\end{figure}

In the presence of new physics from Eq.~\eqref{eq:lag}, the lepton thermal masses would get modified.  Nevertheless,  since  neutrinogenesis  typically culminates at $T< m_\phi$, we expect the thermal correction from    Eq.~\eqref{eq:lag} would be suppressed by the scalar mass.  To see this, let us calculate the contribution from Eq.~\eqref{eq:lag} to the real coefficients $\text{Re}a_{}, \text{Re}b_{}$, which is determined by the real part of the one-loop lepton self-energy diagram, as shown in Fig.~\ref{fig:tildem-NP}. Following Refs.~\cite{Carrington:1997sq,Li:2023ewv}, we arrive at 
\begin{align}\label{eq:ReSigma}
\text{Re}\Sigma_{\ell_i}^{T}(p)=2\pi |y_i|^2 \int\frac{d^4q}{(2\pi)^4}\left\{\frac{\delta[q'^2-m_\phi^2]}{q^2}f_\phi^{\rm eq}(|q'_0|)- \frac{\delta(q^2)}{q'^2-m_\phi^2}f_{\nu_R}^{\rm eq}(|q_0|)\right\} P_R\slashed{q}P_L\,,
\end{align}
where $|y_i|^2\equiv \sum_\alpha |y_{i\alpha}|^2$, $q'=q-p$, and we have used the thermal scalar distribution for simplicity. The above integral can be further simplified by  taking the transformation $q\to -q+p$ in the scalar term such that the distribution functions depend only on $|q_0|$. 

For soft leptons with  $|p_0|\sim |\vec p| \sim |y_i| T$,  the   amplitude given in Eq.~\eqref{eq:ReSigma} has a more complicated structure than the usual situation where no vacuum mass appears, as now it introduces a mass scale beyond the HTL treatment.  Nevertheless, forbidden neutrinogenesis occurs at $T<m_\phi$, which allows us to estimate the   modification from Eq.~\eqref{eq:ReSigma} to the lepton thermal mass by treating $T/m_\phi$ as a small parameter. Numerically, this treatment is    justified even  if the $q$-momentum integration in Eq.~\eqref{eq:ReSigma} is extended to     $p\to \infty$, since the   distribution functions $f_{\nu_R}^{\rm eq}$ and $f_\phi^{\rm eq}$ ensure that  the integral  at $q\gg T$ would be  exponentially suppressed.   

With  the general expression for $\text{Re}a, \text{Re}b$~\cite{Li:2023ewv}:
\begin{align}\label{eq:aL}
	\text{Re}a_i	&=\frac{1}{2|\vec p|^{2}}\left(\text{Tr}[\slashed{p}\text{Re}\Sigma_{\ell_i}^{T}]-p_{0}\text{Tr}[\slashed{u}\text{Re}\Sigma_{\ell_i}^{T}]\right),
	\\[0.2cm]
	\text{Re}b_i	&=-\frac{1}{2|\vec p|^{2}}\left(p_{0}\text{Tr}[\slashed{p}\text{Re}\Sigma_{\ell_i}^{T}]-(p_0^2-|\vec p|^2)\text{Tr}[\slashed{u}\text{Re}\Sigma_{\ell_i}^{T}]\right),\label{eq:bL}
\end{align}
it is straightforward to  derive $	\text{Re}a_i, \text{Re}b_i$  at   leading order of $T/m_\phi$. We have
\begin{align}\label{eq:aL2}
	\text{Re}a_i	&=\frac{|y_{i}|^2}{|\vec p|^{2}} \left[ R_{T/m_\phi,1}(|\vec p|^2+3p_0^2) +R_{T/m_\phi, 2}\,p_0^2\right]\,,
	\\[0.2cm]
	\text{Re}b_i	&=-\frac{|y_{i}|^2}{|\vec p|} \left[ R_{T/m_\phi,1}(|\vec p|^2+3p_0^2)\frac{p_0}{|\vec p|} +R_{T/m_\phi, 2}(p_0^2-|\vec p|^2) \frac{p_0}{|\vec p|}\right],\label{eq:bL2}
\end{align}
where $R_{T/m_\phi,1}$ and $R_{T/m_\phi,2}$ denote the  $T/m_\phi$ functions at leading order:
\begin{align}
R_{T/m_\phi,1}&=-\frac{1}{\sqrt{2\pi^3}}\left(\frac{T}{m_\phi}\right)^{5/2}e^{-m_\phi/T}-\frac{7\pi^2}{360}\left(\frac{T}{m_\phi}\right)^4\,,
\\[0.2cm]
R_{T/m_\phi,2}&=-\frac{1}{\sqrt{8\pi^3}}\left(\frac{T}{m_\phi}\right)^{3/2}e^{-m_\phi/T}+\frac{7\pi^2}{120}\left(\frac{T}{m_\phi}\right)^4\,.
\end{align}
To see how lepton thermal masses are modified, it is instructive to make a comparison between   Eqs.~\eqref{eq:aL2}-\eqref{eq:bL2} and Eqs.~\eqref{eq:aRi}-\eqref{eq:bRi}. We can infer from Eq.~\eqref{eq:p0-p} that the modified  dispersion relation  in the regime   $p_0^2\sim |\vec p|^2\sim  |y_{i}|^2 T^2$ yields
\begin{align}
p_0-|\vec p|\simeq  \frac{|y_{i}|^4  T^2}{|\vec p|} \left(R_{T/m_\phi,1}+R_{T/m_\phi,2}\right)\,.	\label{eq:tildem_y}
\end{align}
 Given that  $|y_i|=\mathcal{O}(0.01)$ is predicted  in the case of one nonthermal $\nu_R$,  we see that the thermal mass correction from neutrino Yukawa couplings  would give 
 \begin{align}
 	\frac{\tilde{m}^2}{T^2}=\mathcal{O}(10^{-8}) \left(R_{T/m_\phi,1}+R_{T/m_\phi,2}\right),
 \end{align}
which is    smaller than the contribution from the muon Yukawa coupling $y_\mu^2=\mathcal{O}(10^{-7})$ at $T<m_\phi$.  Therefore, if forbidden neutrinogenesis culminates at $T<m_\phi$, as confirmed by Fig.~\ref{fig:Y1}, the maximally resonant enhancement from   quasi-degenerate thermal masses, i.e.,  taking $j=\mu, i=e$ in Eq.~\eqref{eq:S_CP_fin-2},  will not be violated by  a   neutrino Yukawa coupling at $\mathcal{O}(0.01)$,

Finally, we would like to comment on the thermal width of soft leptons. 
The zero-width approximation, $\text{Im}\Delta_\pm\to 0$,  has been used to calculate the CP-violating source, which leads to the on-shell resummed propagators:
\begin{align}\label{eq:tildeS^R-onshell}
	i\slashed{{S}}^{R}_{i}(p)\Big|_{\rm onshell}&\approx \pi\text{sign}(p_{0}) \delta(p^2-2\tilde{m}_i^2)\slashed{p}\,,
	\\[0.2cm]
	i\slashed{{S}}_i^{<}(p)\Big|_{\rm onshell}&\approx -2\pi\text{sign}(p_0) f(p_0)\delta(p^2-2\tilde{m}_i^2)\slashed{p}\,,\label{eq:tildeS^<-onshell}
	\\[0.2cm]
	i\slashed{{S}}_i^{>}(p)\Big|_{\rm onshell}&\approx 2\pi\text{sign}(p_0)\left[1-f(p_0)\right]\delta(p^2-2\tilde{m}_i^2)\slashed{p}\,,\label{eq:tildeS^>-onshell}
\end{align}
  where the common pole of   $\slashed{{S}}^R$ and $\slashed{{S}}^\lessgtr$ is  determined by $\text{Re}\Delta_{\pm}=0$, with
\begin{align}\label{eq:pole4}
	p_0= \mp \left[|\vec p|+\frac{\tilde{m}^2}{|\vec p|}-\frac{\tilde{m}^4}{2|\vec p|^3}\log\left(\frac{2|\vec p|^2}{\tilde{m}^2}\right)+\mathcal{O}(\tilde{m}^6)\right].
\end{align}
 As elaborated    in Ref.~\cite{Kanemura:2024dqv},  
including $\text{Im}\Delta_\pm$ at    next-to-leading order of  gauge couplings would give rise to $\text{Im}\Delta_\pm = \kappa (3g_2^2+g_1^2) \tilde{m}^2/|\vec{p}|$  with $\kappa\ll 1$, which  protects   the stability of the resonant enhancement  from finite thermal width near the pole.

\section{CP-violating rate from  three nonthermal neutrinos}\label{append:SCP_EWscalar}
This appendix collects the derivation of the CP-violating rate in the case of   three nonthermal neutrinos, as given in Eq.~\eqref{eq:S_CP_fin} and Eq.~\eqref{eq:statistics}. 

Based on Eq.~\eqref{eq:KB}, we can write down  the    CP-violating source $\mathcal{S}_{\rm CP}$   as
\begin{align}\label{eq:S_CP-2}
	\mathcal{S}_{\rm CP}(p)=\frac{1}{2}\int_p (2\pi)\delta(p^2) \text{Tr}[(\slashed{K}_1+\slashed{K}_2) P_R\slashed{p}P_L]\,,
\end{align}
where
\begin{align}
	\slashed{K}_1&\equiv \theta(-p_{0})i\slashed{\Sigma}_{\nu_\alpha}^>- \theta(p_{0})i\slashed{\Sigma}_{\nu_\alpha}^<
	=-2\int_{p_\ell} \int_{p_\phi} (2\pi)^4\delta^4(p_{}-p_\ell+p_\phi) \sum_{i=1}^{5}\mathcal{I}_i\,,\label{eq:K1}
	\\[0.2cm]
	\slashed{K}_2&\equiv (i\slashed{\Sigma}_{\nu_\alpha}^>-i\slashed{\Sigma}_{\nu_\alpha}^<)\mathcal{F}_\alpha=-2 \mathcal{F}_\alpha\int_{p_\ell} \int_{p_\phi} (2\pi)^4\delta^4(p_{}-p_\ell+p_\phi)  \sum_{i=1}^{5}\mathcal{J}_i\,,\label{eq:K2}
\end{align}
with      $\mathcal{F}_\alpha$  a  neutrino-distribution dependent function: $\mathcal{F}_\alpha\equiv \theta(p_0)f_\alpha(p_0)+\theta(-p_0)\bar f_\alpha(-p_0)$.
The $\slashed{K}_1$  contribution   is independent of the  neutrino distribution functions $f_\alpha$. We will mostly follow the approach presented in Ref.~\cite{Kanemura:2024dqv} to calculate  the $\slashed{K}_1$   term, and then we extend the approach to determine  the $\slashed{K}_2$ contribution that depends on $f_\alpha, \bar f_\alpha$. 

Let us first consider the $\slashed{K}_1$  term.  
Functions  $\mathcal{I}_i$ in $\slashed{K}_1$    result  from the difference of Eq.~\eqref{eq:tildeS<Lij} and Eq.~\eqref{eq:tildeS>Lij}, and are given by 
\begin{align}
	\mathcal{I}_1&=i\slashed{{S}}^R_{\ell_i} \left[y_4^*\theta(-p_{ 0})e^{p_{\ell 0}/T}iG_\phi^<(-i\slashed{\Sigma}^T_{\ell})+y_4\theta(p_{0})iG_\phi^>(-i\slashed{\Sigma}^T_{\ell})\right]i\slashed{{S}}^<_{\ell_j}\,,
	\\[0.2cm]
	\mathcal{I}_2&=-i\slashed{{S}}^R_{\ell_i}  \left[y_4^*\theta(-p_{0})iG_\phi^< (-i\slashed{\Sigma}^>_{\ell})-y_4\theta(p_{0})iG_\phi^>(-i\slashed{\Sigma}^<_{\ell})\right] i\slashed{{S}}^R_{\ell_j}\,,\label{eq:I2}
	\\[0.2cm]
	\mathcal{I}_3&
	=-i\slashed{{S}}^R_{\ell_i}\left[y_4\theta(-p_{0})iG_\phi^<(-i\slashed{\Sigma}^>_{\ell})-y_4^*\theta(p_{0})iG_\phi^>(-i\slashed{\Sigma}^<_{\ell})\right] i\slashed{{S}}^<_{\ell_j}\,,\label{eq:I3}
	\\[0.2cm]
	\mathcal{I}_4&=-i\slashed{{S}}^R_{\ell_i}\left[y_4^*\theta(-p_{0})e^{p_{\ell 0}/T}iG_\phi^<(-i\slashed{\Sigma}^>_{\ell})-y_4\theta(p_{0})e^{p_{\ell 0}/T} iG_\phi^>(-i\slashed{\Sigma}^<_{\ell})\right]i\slashed{{S}}^<_{\ell_j}\,,
	\\[0.2cm]
	\mathcal{I}_5&= -i\slashed{{S}}^R_{\ell_i}\left[y_4 \theta(-p_{0})e^{p_{\ell 0}/T}iG_\phi^<(-i\slashed{\Sigma}^{\bar T}_{\ell})+y_4^*\theta(p_{0})iG_\phi^>(-i\slashed{\Sigma}^{\bar T}_{\ell})\right] i\slashed{{S}}^<_{\ell_j}\,.\label{eq:I5}
\end{align}
The appearance of   $e^{p_{\ell 0}/T}$ in $\mathcal{I}_1,\mathcal{I}_4$ and $\mathcal{I}_5$ arises from the KMS relation
\begin{align}
	\slashed{{S}}^>_{\ell}(p)=-e^{p_{0}/T}\slashed{S}_\ell^<(p)\,, 
\end{align}
which is valid by  neglecting   small chemical potentials  in thermal lepton distribution functions. 
The detailed calculations of each  $\mathcal{I}_i$ can be found in Ref.~\cite{Kanemura:2024dqv}.  The CP-violating source $\mathcal{S}_{\rm CP}$ from the $\slashed{K}_1$ term  gives
\begin{align}\label{eq:S_CP-4}
	S_{\rm CP}^{\slashed{K}_1}(p)=-\frac{(2\pi)^2 \text{Im}(y_4)m_\phi^4}{\tilde{m}_j^2-\tilde{m}_i^2}&\int_{p_i} \tilde{\delta}^4 \left(\sum p_i\right)\delta(p^2)\theta(p_0)\theta(-p_{\ell 0})\delta(p_\ell^2-2\tilde{m}_j^2) iG_\phi^>\mathcal{TC}\,,
\end{align}
where $p_i=p, p_\ell, p_\phi$, $\tilde{\delta}^4\left(\sum p_i \right)\equiv (2\pi)^4\delta^4(p_{}-p_\ell+p_\phi)$ dictates energy-momentum conservation, $\sqrt{2}\tilde{m}_j$ corresponds to the effective thermal mass of lepton flavor $j$~\cite{Weldon:1982bn,Kiessig:2010pr,Li:2023ewv} with $\tilde{m}$ given by Eq.~\eqref{eq:thermalmass}, and the \textit{thermal-criterion} function $\mathcal{TC}$ reads
\begin{align}\label{eq:TC}
\mathcal{TC}\equiv \left(i\Sigma_{\ell}^>-i\Sigma_{\ell}^<\right) f^{\rm eq}_{\ell}(p_{\ell 0})+ i\Sigma_{\ell}^<\,.
\end{align}
This  function provides a clear criterion to check whether the CP-violating source is generated in thermal equilibrium.  If both $\phi$ and $\nu_\beta (\nu_\beta\neq \nu_\alpha)$ in the inner loop are in full thermal equilibrium, the KMS relation holds:
\begin{align}\label{eq:Sigma_ell_KMS}
	\Sigma_\ell^>(p_\ell)=-e^{p_{\ell 0}/T}\Sigma_{\ell}^<(p_\ell)\,,
\end{align}
and  $\mathcal{TC}$ vanishes:
\begin{align}
\mathcal{TC}^{\rm eq}=\left[i\Sigma_{\ell}^{> \,\rm eq}(p_{\ell})- i\Sigma_{\ell}^{<\, \rm eq}(p_{\ell})\right] f^{\rm eq}_{\ell}(p_{\ell 0})+ i\Sigma_{\ell}^{<\,\rm eq}(p_{\ell})=0\,.\label{eq:KMS}
\end{align} 
It indicates that taking a nonthermal neutrino flavor $\alpha (\neq \beta)$ is not sufficient to induce a nonzero CP-violating source, and 
there must be additional nonthermal conditions provided by the inner-loop particles. 

 The quartic scalar mass appearing in Eq.~\eqref{eq:S_CP-4} comes from the Dirac  trace:
\begin{align}\label{eq:Dirac_trace}
	\text{Tr}[P_L\slashed{p}_\ell P_R\slashed{q}P_L \slashed{p}_{\ell}P_R\slashed{p}]=4(p\cdot p_\ell)(q\cdot p_\ell)-2p_\ell^2(p\cdot q)\approx m_\phi^4\,,
\end{align}
where $m_\phi$ in the final approximation should be  taken by the vacuum mass. To see this,
recall that the outer-loop scalar, like the soft leptons, should    be also resummed. Due to the spin-0 nature, the scalar thermal mass can be simply added to the vacuum mass,
\begin{align}\label{eq:thermal_scalar_mass}
	m_\phi^2&=\mu_2^2+m_{\phi, T}^2\,,
\\[0.2cm]
m_{\phi, T}^2&=\left(\frac{3}{16}g_2^2+\frac{1}{16}g_1^2\right)T^2\,,
\end{align}
where we only include the corrections from gauge interactions~\cite{Cline:1995dg}. 
The thermal scalar mass $m_{\phi, T}$ will cancel the lepton thermal mass $\sqrt{2}\tilde{m}$ in the 4-momentum product $p\cdot p_\ell$ and $q\cdot p_\ell$. On the other hand, forbidden neutrinogenesis is an IR-dominated process, which culminates at $T=\mathcal{O}(m_\phi)$. Given this, we neglect the   term proportional to $p_\ell^2=2\tilde{m}^2$.

To proceed with Eq.~\eqref{eq:S_CP-4} in the case of three nonthermal $\nu_R$, we define  $f_\beta=f_\beta^{\rm eq}+\delta f_\beta$ ($\delta f_\beta <0$) for right-handed neutrinos in the inner loop. Then  the propagators and self-energy amplitudes can be written  as
\begin{align}
	i\slashed{S}^{ab}_{\nu_\beta} (p)&=i\slashed{S}^{ab\, \rm eq}_{\nu_\beta}(p)+i \delta \slashed{S}^{}_{\nu_\beta}(p)\,,
	\\[0.2cm]
	i\slashed{\Sigma}_\ell^{ab}(p)&=i\slashed{\Sigma}_{\ell}^{ab\,\rm eq}(p)+i\delta \slashed{\Sigma}_\ell^{ab}(p)\,,\label{eq:S-Sigma_per}
\end{align}
with $a,b=\pm$, $i \delta \slashed S_{\nu_\beta}(p)=-2\pi \delta(p^2)   \slashed{p}  \delta f_{\beta}(|p_0|)$, and 
\begin{align}
	i \slashed{\Sigma}_{\ell}^{ab\,\rm eq}(p_\ell)&= \int_{q,q_\phi} (2\pi)^4 \delta^4(q-q_\phi-p_\ell) P_R i\slashed{S}_{\nu_\beta}^{ab\,\rm eq}P_L iG_\phi^{ba\,\rm eq}\,,
	\\[0.2cm]
	i\delta \slashed{\Sigma}_{\ell}^{ab}(p_\ell)&= \int_{q,q_\phi} (2\pi)^4 \delta^4(q-q_\phi-p_\ell) P_Ri\delta \slashed{S}_{\nu_\beta}^{}P_L i G^{ba\,\rm eq}_\phi\,,
\end{align}
with $q,q_\phi$ the inner-loop momenta of $\nu_\beta$ and $\phi$, respectively. Note that we have neglected the quadratic term $\delta\slashed{S}_{\nu_\beta}\delta G_\phi$, and also the term $\slashed{S}^{\rm eq}_{\nu_\beta}\delta G_\phi$. The reason will become  clear  after we compare the results between the cases of three nonthermal $\nu_R$ (section~\ref{sec:Y3}) and one nonthermal $\nu_R$ (section~\ref{sec:Y1}).  In both cases, $\delta f_\phi=f_\phi -f_\phi^{\rm eq}$ is expected to be small due to the quasi-thermal nature of the neutrinophilic scalar. Nevertheless, as mentioned in section~\ref{sec:Y1}, a small $\delta f_\phi$ can be compensated for by larger Yukawa couplings from the two heavier $\nu_R$. For the case of three nonthermal $\nu_R$, all the neutrino Yukawa couplings are small and hence no coupling enhancement to compensate for the  $\delta f_\phi$  suppression. 

Substituting Eq.~\eqref{eq:S-Sigma_per} into Eq.~\eqref{eq:S_CP-4} and using Eq.~\eqref{eq:KMS},  we can reduce the   thermal-criterion function $\mathcal{TC}$ to 
\begin{align}\label{eq:bracket_fin}
	\mathcal{TC}=-\frac{1}{8\pi p_\ell}\int^\infty_{m_\phi^2/(4p_\ell)} dq \delta f_\beta(q)\left[f^{\rm eq}_{\ell}(p_\ell)+f^{\rm eq}_\phi(q+p_\ell)\right],
\end{align}
where $q\equiv |\vec q|, p_\ell\equiv |\vec p_\ell|$, and $q_0>0, q_{\phi 0}=q_0-p_{\ell 0}>0$ were used. The lower integration limit of $q$ comes from the angular integration with $\delta(-m_\phi^2+2q  p_\ell+2q p_\ell \cos\theta)$, and  we have  used
$f_{\ell}(p_{\ell 0})=1-f_{\ell}(-p_{\ell 0})$,
for $p_{\ell 0}=-\omega_j$ with the  approximation $ \omega_j\approx  p_\ell$.

In Eq.~\eqref{eq:S_CP-4}, we can integrate over $d^4 p$ via $\delta^4(p-p_\ell+p_\phi)$, $dp_{\ell 0}$ via $\delta(p^2_{\ell}-2\tilde{m}_j^2)$, and $dp_{\phi 0}$ via $\delta(p_\phi^2-m_\phi^2)$ from the scalar Wightman function $G_\phi^>$. Finally, we   arrive  at  the CP-violating source $\mathcal{S}_{\rm CP}$ from the $\slashed{K}_1$ term: 
\begin{align}\label{eq:SCP_K1}
	S_{\rm CP}^{\slashed{K}_1}=\frac{\text{Im}(y_4)m_\phi^4}{256\pi^4 (\tilde{m}_j^2-\tilde{m}_i^2)}\int^\infty_0 \frac{dp_\ell}{p_\ell}\int^\infty_{\frac{m_\phi^2}{4p_\ell}+p_\ell}  f^{\rm eq}_\phi(E_\phi)dE_\phi \int^\infty_{\frac{m_\phi^2}{4p_\ell}} dq \delta f_\beta(q)\left(f^{\rm eq}_{\phi}+f^{\rm eq}_\ell\right),
\end{align}
where $(f^{\rm eq}_\phi+f^{\rm eq}_\ell)\equiv f^{\rm eq}_{\phi}(q+p_\ell)+f^{\rm eq}_\ell(p_\ell)$, and  the lower integration limit  of $E_\phi$ comes from the angular integration with $\delta(m_\phi^2-2E_\phi p_\ell+2p_\ell p_\phi \cos\theta)$.  
Next, let us turn to evaluate the $\slashed{K}_2$ term. 
With the same simplification in calculating  $\mathcal{I}_i$, it is straightforward to obtain the five $\mathcal{J}_i$ functions of  $\slashed{K}_2$ in Eq.~\eqref{eq:K2}:
\begin{align}
	\mathcal{J}_1&=i\slashed{{S}}^R_{\ell_i} \left[y_4^*e^{p_{\ell 0}/T}iG_\phi^<(-i\slashed{\Sigma}^T_{\ell})+y_4 iG_\phi^>(-i\slashed{\Sigma}^T_{\ell})\right]i\slashed{{S}}^<_{\ell_j}\,,\label{eq:J1}
	\\[0.15cm]
	\mathcal{J}_2&=-i\slashed{{S}}^R_{\ell_i}  \left[y_4^* iG_\phi^< (-i\slashed{\Sigma}^>_{\ell})-y_4 iG_\phi^>(-i\slashed{\Sigma}^<_{\ell})\right] i\slashed{{S}}^R_{\ell_j}\,,\label{eq:J2}
	\\[0.15cm]
	\mathcal{J}_3&=-i\slashed{{S}}^R_{\ell_i}\left[y_4iG_\phi^<(-i\slashed{\Sigma}^>_{\ell})-y_4^*iG_\phi^>(-i\slashed{\Sigma}^<_{\ell})\right] i\slashed{{S}}^<_{\ell_j}\,,\label{eq:J3}
	\\[0.15cm]
	\mathcal{J}_4&=-i\slashed{{S}}^R_{\ell_i}\left[y_4^*e^{p_{\ell 0}/T}iG_\phi^<(-i\slashed{\Sigma}^>_{\ell})-y_4e^{p_{\ell 0}/T} iG_\phi^>(-i\slashed{\Sigma}^<_{\ell})\right]i\slashed{{S}}^<_{\ell_j}\,,\label{eq:J4}
	\\[0.15cm]
	\mathcal{J}_5&=-i\slashed{{S}}^R_{\ell_i}\left[y_4 e^{p_{\ell 0}/T}iG_\phi^<(-i\slashed{\Sigma}^{\bar T}_{\ell})+y_4^*iG_\phi^>(-i\slashed{\Sigma}^{\bar T}_{\ell})\right] i\slashed{{S}}^<_{\ell_j}\,.\label{eq:J5}
\end{align}
To proceed with the $\slashed{K}_2$ contribution, we neglect the difference $f_\alpha-\bar f_\alpha$,  so that    $\mathcal{F}_\alpha=f_\alpha(|p_0|)$. This approximation amounts to neglecting the washout rate at two-loop order, which is at $\mathcal{O}(y^4)$ and  smaller than $\mathcal{O}(y^2)$  at one-loop order. 
Writing the CP-violating source from the $\slashed{K}_2$ term as 
\begin{align}\label{eq:S_CP-5}
	\mathcal{S}^{\slashed{K}_2}_{\rm CP}(p)=-\int_{p,p_\ell, p_\phi} (2\pi)^5\delta^4(p_{}-p_\ell+p_\phi)\delta(p^2)\mathcal{F}_\alpha \sum_{i=1}^{5}\mathcal{J}_{\text{CP}i}\,,
\end{align}
where $\mathcal{J}_{\text{CP}i}\equiv \text{Tr}[\mathcal{J}_i P_R \slashed{p}P_L]$ 
with $\mathcal{J}_i$ given by Eqs.~\eqref{eq:J1}-\eqref{eq:J5}, we arrive at 
\begin{align}
	\mathcal{J}_{\text{CP}1}&=\frac{2\pi \text{Im}(y_4)m_\phi^4}{\tilde{m}_j^2-\tilde{m}_i^2 } iG_{\phi}^{>}(-i \Sigma^T_{\ell})\text{sign}(p_{\ell 0})  f_\ell(p_{\ell 0})\delta(p_\ell^2-2\tilde{m}_j^2)\,,
	\\[0.15cm]
	\mathcal{J}_{\text{CP}2}&=\frac{ i\pi (y_4^*+y_4) m_\phi^4}{\tilde{m}_j^2-\tilde{m}_i^2}iG_\phi^> (-i\Sigma_\ell^<)\text{sign}(p_{\ell 0}) \delta(p_\ell^2-2\tilde{m}_j^2)\,,\label{eq:J_CP2}
	\\[0.15cm]
	\mathcal{J}_{\text{CP}3}&=\frac{-\pi  [iy_4+2\text{Im}(y_4)f_\ell(p_{\ell 0})]m_\phi^4}{\tilde{m}_j^2-\tilde{m}_i^2}iG_\phi^> (-i\Sigma_\ell^<)\text{sign}(p_{\ell 0}) \delta(p_\ell^2-2\tilde{m}_j^2)=\mathcal{J}_{\text{CP}4}\,,\label{eq:J_CP3}
	\\[0.15cm]
	\mathcal{J}_{\text{CP}5}&=\frac{2\pi \text{Im}(y_4)m_\phi^4}{\tilde{m}_j^2-\tilde{m}_i^2 } iG_{\phi}^{>}(-i \Sigma^{\bar T}_{\ell})\text{sign}(p_{\ell 0})  f_\ell(p_{\ell 0})\delta(p_\ell^2-2\tilde{m}_j^2)\,.
\end{align}
Assembling these $\mathcal{J}_{\text{CP}i}$ functions and using the kinetic condition $p_0p_{\ell 0}<0$ from Dirac $\delta$-functions, we have 
\begin{align}\label{eq:Fasign}
	\text{sign}(p_{\ell 0})\mathcal{F}_\alpha=- [\theta(p_0)\theta(-p_{\ell 0})-\theta(-p_0)\theta(p_{\ell 0})]f_\alpha(|p_0|)\,,
\end{align}
 finally leading us to  arrive at 
\begin{align}\label{eq:SCP_K2}
	\mathcal{S}^{\slashed{K}_2}_{\rm CP}(p)=\frac{(2\pi)^3 \text{Im}(y_4)m_\phi^4}{\tilde{m}_j^2-\tilde{m}_i^2}&\int_{p_i} \tilde{\delta}^4\left(\sum p_{i}\right)\delta(p^2) \theta(p_0)\theta(-p_{\ell 0}) f_\alpha(|p_0|)  \delta_\ell \delta_\phi \mathcal{TC}
\end{align}
where  $p_i=p, p_\ell, p_\phi$, $\tilde{\delta}^4\left(\sum p_i \right)\equiv (2\pi)^4\delta^4(p_{}-p_\ell+p_\phi)$, $\delta_\ell\equiv \delta(p_\ell^2-2\tilde{m}_j^2), \delta_\phi\equiv \delta(p_\phi^2-m_\phi^2)$,  and the thermal-criterion function $\mathcal{TC}$  is given by Eq.~\eqref{eq:bracket_fin}. Note that we have performed momentum reflection $p_i\to -p_i$ in the second term of Eq.~\eqref{eq:Fasign} to obtain Eq.~\eqref{eq:SCP_K2}. Combining Eq.~\eqref{eq:SCP_K1} and Eq.~\eqref{eq:SCP_K2} will lead to the final CP-violating rate in the case of three nonthermal  $\nu_R$, as given in Eq.~\eqref{eq:S_CP_fin} and Eq.~\eqref{eq:statistics}.

\bibliographystyle{JHEP}
\bibliography{Refs}

\end{document}